\documentclass[12pt,twoside]{amsart}

\usepackage[body={16cm,23cm}]{geometry}
\usepackage{cite}
\usepackage{amsmath,amssymb}
\usepackage{amsfonts}
\usepackage[mathscr]{eucal}
\usepackage{epic}
\usepackage{eepic}

\newcommand{\ds}{\displaystyle}
\def\EXP{\textrm{{\large e}}}
\newcommand{\uop}{\mathbf{u}}
\newcommand{\vop}{\mathbf{v}}
\newcommand{\wop}{\mathbf{w}}
\newcommand{\xop}{\mathbf{x}}
\newcommand{\yop}{\mathbf{y}}

\newcommand{\alg}{\mathcal{H}}
\newcommand{\calg}{A}
\newcommand{\qalg}{\mathcal{A}}
\newcommand{\vrt}{\texttt{v}}
\newcommand{\Tau}{\textrm{\large $\tau$}}

\newcommand{\Nop}{\mathbf{h}}
\newcommand{\Rop}{\mathbf{R}}
\def\tvec#1#2{\left(\begin{array}{ll} #1 \\ #2 \end{array}\right)}

\begin{document}

\vspace{2cm}

\title[]{Quantum curve in $q$-oscillator model}%
\author{S. Sergeev}%
\address{Department of Theoretical Physics,
Research School of Physical Sciences and Engineering, Australian
National University, Canberra ACT 0200, Australia}
\email{sergey.sergeev@anu.edu.au}

\thanks{This work was supported by the Australian Research Council}%

\subjclass{37K15}%
\keywords{Quantum integrable system, Tetrahedron equation, Spectral curve, nested Bethe Ansatz}%

\begin{abstract}
A lattice model of interacting q-oscillators, proposed in [V.
Bazhanov, S. Sergeev, arXiv:hep-th/0509181], is the quantum
mechanical integrable model in 2+1 dimensional space-time. Its
layer-to-layer transfer-matrix is a polynomial of two spectral
parameters, it may be regarded in the terms of quantum groups both
as a sum of sl(N) transfer matrices of a chain of length M and as
a sum of sl(M) transfer matrices of a chain of length N for
reducible representations. The aim of this paper is to derive the
Bethe Ansatz equations for the q-oscillator model entirely in the
framework of 2+1 integrability providing the evident rank-size
duality.
\end{abstract}

\maketitle

\section*{Introduction}

The $q$-oscillator lattice model was formulated recently in
\cite{First,letter}. It describes a system of interacting
$q$-oscillators situated in the vertices of two dimensional
lattice, and therefore it is the quantum mechanical system in
$2+1$ dimensional space-time in the same way, as a chain of
interacting particles (or spins) is regarded as a model in $1+1$
dimensional space-time. Formulation of the $q$-oscillator model
provides a definition of a layer-to-layer transfer matrix as a
polynomial of two spectral parameters. This transfer matrix may be
interpreted in the terms of quantum inverse scattering method and
quantum groups, so that both sizes of the two dimensional lattice
may be interpreted as either a length of an effective chain or as
symmetry group's rank. This was called in \cite{First}  the
``rank-size'' duality. The appearance of a complete set of
fundamental transfer matrices for $\mathcal{U}_q(\widehat{sl})$
series is a signal that the layer-to-layer transfer matrix of
$q$-oscillator model is closely related to Bethe Ansatz in the
form of generalized Baxter's ``T-Q'' equations. The subject of
this paper is the derivation of such equations in the framework of
$2+1$ dimensional integrability.

Below in this Introduction we formulate the answer, i.e. we give
an explicit form of ``T-Q'' equations in the terms of a given
layer-to-layer transfer matrix. To do this, we need to repeat the
structure of the layer-to-layer transfer matrix for $q$-oscillator
model in more details.

The $q$-oscillator model describes a system of interacting
$q$-oscillators $\alg_\vrt$,
\begin{equation}\label{q-osc}
\alg_\vrt\ :\ \xop_\vrt\yop_\vrt=1-q^{2+2\Nop_\vrt}\;,\quad
\yop_\vrt\xop_\vrt=1-q^{2\Nop_\vrt}\;,\quad \xop_\vrt
q^{\Nop_\vrt}=q^{\Nop_\vrt+1}\xop_\vrt\;,\quad \yop_\vrt
q^{\Nop_\vrt} = q^{\Nop_\vrt-1}\yop_\vrt\;,
\end{equation}
situated in the vertices $\vrt$ of two dimensional square lattice
of sizes $N\times M$. Index $\vrt$ stands for a coordinate of a
vertex. Oscillators from different vertices commute (what is
called ``locality''), the whole \emph{algebra of observables} is
thus $\alg^{\otimes NM}$, and the vertex index $\vrt$ corresponds
to the number of component of the tensor power. In this paper we
imply mostly the Fock space representation $\mathcal{F}$ of
$q$-oscillators.

The two dimensional lattice may be identified with a layer (or a
section) of three dimensional cubic lattice, further we call it
either the layer or the auxiliary lattice.

Auxiliary matrices $L_{\alpha,\beta}[\alg_\vrt]$, acting in
$\mathbb{C}^2\otimes \mathbb{C}^2\otimes \mathcal{F}_\vrt$, were
introduced in \cite{First}. The \emph{layer} transfer matrix
$\mathbf{T}(u,v)$ may be constructed as a trace of an $2d$ ordered
product of auxiliary matrices $L[\alg_\vrt]$. The transfer matrix
is a polynomial of two spectral parameters,
\begin{equation}\label{transfer-matrix}
\mathbf{T}(u,v)\ = \ \sum_{n=0}^{N}\; \sum_{m=0}^{M} \ u^n v^m\;
\mathbf{t}_{n,m}\ ,
\end{equation}
its coefficients $\mathbf{t}_{n,m}\in\alg^{\otimes NM}$ form a
complete commutative set. Matrices $L[\alg_\vrt]$ depend on some
extra $\mathbb{C}$-valued free parameters, for their generic
values the model is inhomogeneous. The layer transfer matrix
(\ref{transfer-matrix}) may be identically rewritten in two ways,
\begin{equation}\label{dec}
\mathbf{T}(u,v)\ \equiv \ \sum_{n=0}^N \; u^n \;
T^{(sl_N)}_{\omega_n}(v) \ \equiv \ \sum_{m=0}^{M} \; v^m
T^{(sl_M)}_{\omega_m}(u)\;,
\end{equation}
where
\begin{equation}\label{TslN}
T^{(sl_N)}_{\omega_n}(v) \ = \ \sum_{m=0}^M \; v^m
\mathbf{t}_{n,m}
\end{equation}
is the $2d$ transfer matrix for $\mathcal{U}_q(\widehat{sl}_N)$
\emph{chain} of the length $M$, corresponding to the fundamental
representation $\pi_{\omega_n}$ in the auxiliary space (here
$\omega_n$ stand for the fundamental weights of $A_{N-1}$,
$\pi_{\omega_0}$ and $\pi_{\omega_N}$ are two scalar
representations, $T_0^{(sl_N)}$ and $T_N^{(sl_N)}$ may be written
explicitly). The same layer transfer matrix $\mathbf{T}(u,v)$ may
be rewritten as the sum of $\mathcal{U}_q(\widehat{sl}_M)$
transfer matrices
\begin{equation}\label{TslM}
T^{(sl_M)}_{\omega_m}(u) \ = \ \sum_{n=0}^N \; u^n
\mathbf{t}_{n,m}
\end{equation}
for the length $N$ chain (the last part of (\ref{dec})).

The result of this paper is the derivation of the dual Bethe
Ansatz equations for the $q$-oscillator model. They may be
formulated as follows. Let normalized transfer matrices be
\begin{equation}\label{Tau}
\Tau_m^{(sl_M)}(u) \;=\; T_{\omega_m}^{(sl_M)}(-(-q)^mu)\;, \qquad
\Tau_n^{(sl_N)}(v) \;=\; T_{\omega_n}^{(sl_n)}(-(-q)^nv)\;.
\end{equation}
And let $\mathbb{C}$-numerical parameters of $q$-oscillator
lattice are inhomogeneous enough.

Then ``T-Q'' equation for $sl_M$ is
\begin{equation}\label{BAE-slM}
\sum_{m=0}^M (-v)^m \ \Tau_m^{(sl_M)}(u) \ Q(q^{2m}u)   \ = \ 0\;.
\end{equation}
The statement is that if $\mathbf{t}_{n,m}$ take their
eigenvalues, then there exist $M$ special values $v_1,\dots,v_M$
of $v$, such that corresponding $Q_1(u),\dots, Q_M(u)$ in
(\ref{BAE-slM}) are polynomials\footnote{parameter $v$ in the
``$u$-shift'' equation (\ref{BAE-slM}) is irrelevant since a
re-scaling $Q(u)\to u^\nu Q(u)$ changes it.}. Degrees of the
polynomials are uniquely defined by certain occupation numbers of
oscillators.

In its turn, equivalent ``T-Q'' equation for $sl_N$ is
\begin{equation}\label{BAE-slN}
\sum_{n=0}^N  (-u)^n \ \Tau_n^{(sl_N)}(v) \ \overline{Q}(q^{2n}v)
\ = \ 0\;.
\end{equation}
If $\mathbf{t}_{n,m}$ take their eigenvalues, then there exist $N$
special values $u_1,\dots, u_N$ of $u$, such that corresponding
$\overline{Q}_1(v),\dots,\overline{Q}_N(v)$ in (\ref{BAE-slN}) are
polynomials. All the other forms of nested Bethe Ansatz equations
follow from (\ref{BAE-slM}) or (\ref{BAE-slN}).

Polynomials $Q(u)$ and $\overline{Q}(v)$ may be denoted in the
quantum mechanical way as ``wave functions'' of states $\langle
Q|$ and $|\overline{Q}\rangle$:
\begin{equation}
Q(u)\;=\;\langle Q|u\rangle\quad \textrm{and}\quad
\overline{Q}(v)\;=\;\langle v|\overline{Q}\rangle\;,
\end{equation}
where $|u\rangle$ and $\langle v|$ serve the simple Weyl algebra
$\ds \mathcal{W}$: $\uop\;\vop\;=\;q^{2}\; \vop\;\uop$,
\begin{equation}\label{Weyl}
\begin{array}{l}
\ds \uop|u\rangle \;=\; |u\rangle u\;,\quad \vop|u\rangle \;=\;
|q^2u\rangle v\;,\\
\textrm{and}\\
\ds \langle v|\uop\;=\; u\langle q^2v|\;, \quad \langle
v|\vop\;=\; v\langle v|\;.
\end{array}
\end{equation}
Let
\begin{equation}\label{J}
J(\uop,\vop)\;=\;\sum_{n=0}^N\sum_{m=0}^M (-q)^{-nm} (-\uop)^n
(-\vop)^m t_{n,m}\;.
\end{equation}
Then (\ref{BAE-slM}) and (\ref{BAE-slN}) are correspondingly
\begin{equation}\label{qc}
\langle Q|\ J(\uop,\vop) \ |u\rangle \;=\; \langle v| \
J(\uop,\vop) \ |\overline{Q}\rangle\;=\;0\;.
\end{equation}

The formulation of the $q$-oscillator model and definition of
$\mathbf{T}(u,v)$ are locally $3d$ invariant, the quantum group
interpretation (\ref{dec}) is the secondary one. In this paper we
will derive (\ref{J}) without any quantum group technique.

To explain our method, we need to comment a little the classical
limit.

In the classical limit $q\to 1$, the local $q$-oscillator
generators become the classical dynamical variables, the
$q$-oscillator model becomes a model of classical mechanics,
quantum evolution operators become Baecklund transformations for
the dynamical variables. In particular, $T(u,v)$ may be understood
as a partition function of a completely inhomogeneous free fermion
six-vertex model on the square lattice (but it should not be
regarded as a model of statistical mechanics). In its turn,
$J(u,v)$ becomes a free fermion determinant (the sign
$(-)^{nm+n+m}$ counts the number of fermionic loops). There exists
a well known formula in the theory of two dimensional free fermion
models, relating $T$ and $J$:
\begin{equation}
T(u,v)\;=\;\frac{1}{2} \left( J(-u,v) + J(u,-v) + J(-u,-v) -
J(u,v)\right)\;.
\end{equation}
In the classical limit, equation $J(u,v)=0$ defines the spectral
curve. Dynamical variables may be expressed in the terms of
$\theta$-functions on the Jacobian of the spectral curve. The
sequence of Baecklund transformations, which is the ``discrete
time'' in the classical model, is a sequence of linear shifts of a
point on the Jacobian. The classical model was formulated and
solved by I. G. Korepanov \cite{Korepanov}.

Classical integrability is based on an auxiliary linear problem.
Equation $J(u,v)=0$ is the condition of the existence of a
solution of the linear problem. Our point is that in quantum
$q\neq 1$ case, the linear problem is still the basic concept of
the solvability. Quantum $\mathbf{J}(\uop,\vop)$ is a well defined
determinant of an operator-valued matrix, and
$\mathbf{J}(\uop,\vop)|\Psi\rangle = 0 $ is again the condition of
the existence of a solution of a quantum linear problem. The
polynomial structure of e.g. $\langle v|\Psi\rangle$ follows from
a more detailed consideration of the quantum linear problem in a
special basis of diagonal ``quantum Baker-Akhiezer function''
(related to a quantum separation of variables).

The structure of the paper is the following. In the first section
we recall briefly some basic notions of the classical model
\cite{Korepanov}: the linear problem, spectral curve and details
of the combinatorial representation of the spectral curve. In the
second section we repeat the definition of the quantum model and
its integrability \cite{First,letter}. In particular, our
definition of the spectral parameters differs from that of
\cite{First}. Quantum linear problem, derivation of (\ref{J}) and
properties of various forms of (\ref{qc}) are given in the third
section. The fourth section includes an example.

\section{The Korepanov model}

We start with a short review of the integrable model of classical
mechanics in discrete $2+1$ dimensional space-time
\cite{Korepanov}. The main purpose of this section is to recall
the relation between Korepanov's linear problem, spectral
determinant and partition function for free fermion model. Another
aim is to fix several useful definition and notations.

\subsection{Linear problem.}

Consider a two dimensional lattice formed by the intersection of
straight lines enumerated by the Greek letters. Let the vertices
of the lattice are enumerated in some way.

\begin{figure}[ht]
\begin{center}
\setlength{\unitlength}{0.25mm}
\begin{picture}(450,200)
\put(0,0){\begin{picture}(200,200)
\thinlines\drawline[-30](30,100)(170,100)
\drawline[-30](40,70)(160,130) \Thicklines \path(100,85)(100,115)
\put(5,97){\scriptsize $\psi_\beta$}\put(23,57){\scriptsize
$\psi_\alpha$} \put(165,135){\scriptsize $\psi_\alpha'$}
\put(175,95){\scriptsize $\psi_\beta'$} \put(95,70){\scriptsize
$\calg_\vrt$}
\end{picture}}
\put(270,95){$\ds \left\{\begin{array}{ll} \ds \psi_\alpha' \; =
\; a_\vrt^{}\psi_\alpha^{} \; + \; b_\vrt^{}\psi_\beta^{}\;,\\ \ds
\psi_\beta' \; = \; c_\vrt^{}\psi_\alpha^{} \; + \;
d_\vrt^{}\psi_\beta^{}\;.\end{array}\right.$}
\put(220,95){$\Leftrightarrow$}
\end{picture}
\end{center}
\caption{Vertex $\vrt$ is formed by intersection of $\alpha$- and
$\beta$-lines of auxiliary lattice. Vertex linear problem is the
pair of relations binding four edge variables.} \label{fig-LP}
\end{figure}
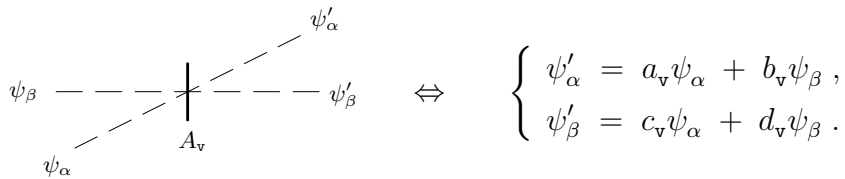

Consider a particular vertex with a number $\vrt$ formed by the
intersection of lines $\alpha$ and $\beta$, as it is shown in Fig.
\ref{fig-LP}. It was mentioned in the Introduction, an auxiliary
lattice is a section of three dimensional lattice, the vertices on
the auxiliary lattice are equivalent to the edges of the
three-dimensional one. In Fig. \ref{fig-LP}, the dashed lines are
the lines of auxiliary lattice, while the solid sprout from the
vertex $\vrt$ is the edge of the three dimensional lattice.

Let four free $\mathbb{C}$-valued variables
\begin{equation}\label{abcd}
\calg_\vrt \; = \; (a_\vrt,b_\vrt,c_\vrt,d_\vrt)
\end{equation}
are associated with vertex $\vrt$. In addition, let
$\mathbb{C}$-valued variables $\psi_\alpha$ and $\psi_\beta$ are
associated with the ingoing edges, and $\mathbb{C}$-valued
variables $\psi_\alpha'$ and $\psi_\beta'$ are associated with the
outgoing edges, as it is shown in Fig. \ref{fig-LP} (a certain
orientation of auxiliary lines is implied).
The local linear problem is a pair of linear relations binding the
edge variables. Its standard form, the right hand side of Fig.
\ref{fig-LP} in matrix notations, is
\begin{equation}\label{X-kor}
\tvec{\psi_\alpha'}{\psi_\beta'}\;=\; X[\calg_\vrt]
\tvec{\psi_\alpha^{}}{\psi_\beta^{}},\;\;\;\textrm{where}\;\;\;
X[\calg_\vrt]\;\stackrel{\textrm{def}}{=}\;\left(\begin{array}{cc} a_\vrt & b_\vrt \\
c_\vrt & d_\vrt
\end{array}\right)\;.
\end{equation}

\subsection{Korepanov's equation}

\begin{figure}[ht]
\begin{center}
\setlength{\unitlength}{0.20mm}
\begin{picture}(750,340)
\put(50,20) {\begin{picture}(300,300)
\Thicklines
 \path(100,50)(200,250)\path(200,0)(100,300)\path(0,250)(300,50)
 \put(195,-20){\scriptsize $\calg_2$}\put(95,310){\scriptsize $\calg_2'$}
 \put(90,30){\scriptsize $\calg_3$}\put(200,260){\scriptsize $\calg_3'$}
 \put(300,30){\scriptsize $\calg_1$}\put(-10,260){\scriptsize $\calg_1'$}
\thinlines
 \drawline[-30](87,69)(213,6)
 \put(67,77){\scriptsize $\psi_\gamma^{}$}
 \put(215,0){\scriptsize $\psi_\gamma'$}
 \drawline[-30](69,60)(321,60)
 \put(45,55){\scriptsize $\psi_\beta^{}$}
 \put(325,55){\scriptsize $\psi_\beta'$}
 \drawline[-30](177,6)(303,69)
 \put(167,-4){\scriptsize $\psi_\alpha^{}$}
 \put(310,75){\scriptsize $\psi_\alpha'$}
\end{picture}}
\put(400,20) {\begin{picture}(300,300)
\Thicklines
 \path(100,50)(200,250)\path(200,0)(100,300)\path(0,250)(300,50)
 \put(195,-20){\scriptsize $\calg_2$}\put(95,310){\scriptsize $\calg_2'$}
 \put(90,30){\scriptsize $\calg_3$}\put(200,260){\scriptsize $\calg_3'$}
 \put(300,30){\scriptsize $\calg_1$}\put(-10,260){\scriptsize $\calg_1'$}
\thinlines
 \drawline[-30](-3,231)(123,294)
 \put(-10,216){\scriptsize $\psi_\alpha^{}$}
 \put(125,300){\scriptsize $\psi_\alpha'$}
 \drawline[-30](-21,240)(231,240)
 \put(-40,237){\scriptsize $\psi_\beta^{}$}
 \put(240,237){\scriptsize $\psi_\beta'$}
 \drawline[-30](87,294)(213,231)
 \put(67,300){\scriptsize $\psi_\gamma^{}$}
 \put(215,221){\scriptsize $\psi_\gamma'$}
\end{picture}}
\end{picture}
\end{center}
\caption{Left and right hand sides of Korepanov equation.}
\label{fig-palms}
\end{figure}
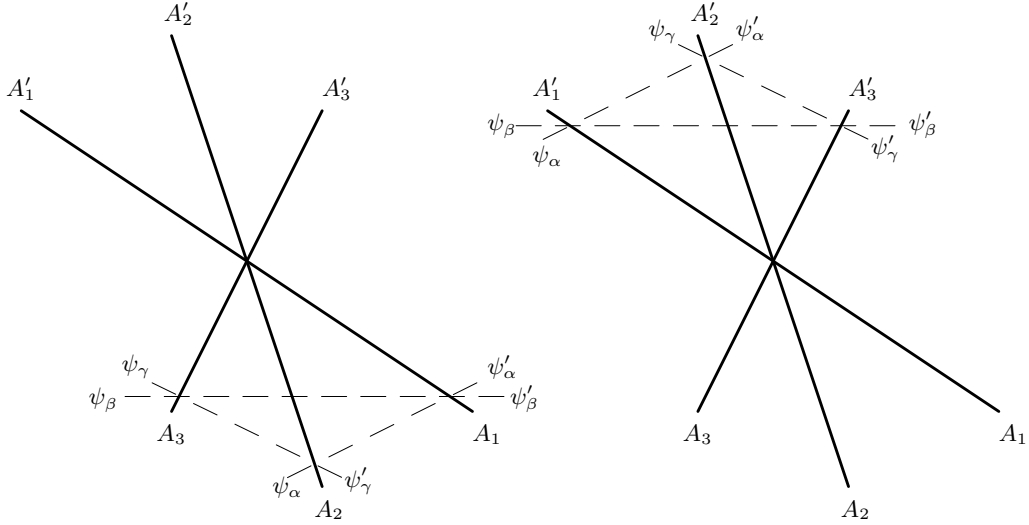

Equations of motion in an integrable model arise as an
associativity condition of its linear problem. To derive their
local form, consider a vertex of three dimensional lattice (not
necessarily the cubic one), and sections of it by two auxiliary
planes, as it is shown in Fig. \ref{fig-palms}. From three
dimensional point of view, the auxiliary linear variables belong
to the faces of 3d lattice, while the dynamical variables
$\calg_\vrt$ belong to the edges of the 3d lattice. Therefore
$\calg_\vrt^{}$ are distinguished from $\calg_\vrt'$, but linear
variables on outer edges $\psi_\alpha^{}$, \dots, $\psi_\gamma'$,
in both top and bottom auxiliary planes are identified.
Consider the bottom plane first. The linear problem rule
(\ref{X-kor}) may be applied three times for excluding internal
edges, as the result one obtains an expression of the ``primed''
linear variables in the terms of ``unprimed'':
\begin{equation}\label{KE-left}
\left(\begin{array}{c} \psi_\alpha' \\ \psi_\beta' \\
\psi_\gamma' \end{array}\right)\;=\;
X_{\alpha,\beta}[\calg_1]\cdot X_{\alpha,\gamma}[\calg_2]\cdot
X_{\beta,\gamma}[\calg_3]\; \cdot\; \left(
\begin{array}{c}
\psi_\alpha \\ \psi_\beta \\ \psi_\gamma\end{array} \right)\;,
\end{equation}
where (cf. the ordering of $\alpha,\beta,\gamma$ in column
vectors)
\begin{equation}\label{imb}
X_{\alpha,\beta}[\calg_1]=\left(\begin{array}{ccc} a_1 & b_1 & 0 \\
c_1 & d_1 & 0 \\ 0 & 0 & 1\end{array}\right)\;,\quad
X_{\alpha,\gamma}[\calg_2]=\left(\begin{array}{ccc} a_2 & 0 & b_2 \\
0 & 1 & 0 \\ c_2 & 0 & d_2 \end{array}\right)\;,\quad
X_{\beta,\gamma}[\calg_3]=\left(\begin{array}{ccc} 1 & 0 & 0 \\ 0 & a_3 & b_3 \\
0 & c_3 & d_3\end{array}\right)
\end{equation}
The top plane of Fig. \ref{fig-palms} may be considered in the
same way,
\begin{equation}\label{KE-right}
\left(\begin{array}{c} \psi_\alpha' \\ \psi_\beta' \\
\psi_\gamma' \end{array}\right)\;=\;
X_{\beta,\gamma}[\calg_3']\cdot X_{\alpha,\gamma}[\calg_2']\cdot
X_{\alpha,\beta}[\calg_1']\; \cdot\; \left(
\begin{array}{c}
\psi_\alpha \\ \psi_\beta \\ \psi_\gamma\end{array} \right)\;,
\end{equation}
where the matrices $X_{\#,\#}$ are given by (\ref{imb}) with
$\calg_\vrt'=(a_\vrt',b_\vrt',c_\vrt',d_\vrt')$.

The associativity condition of linear problems (\ref{KE-left}) and
(\ref{KE-right}) is the Korepanov equation
\begin{equation}\label{KE}
X_{\alpha,\beta}[\calg_1^{}]\cdot
X_{\alpha,\gamma}[\calg_2^{}]\cdot X_{\beta,\gamma}[\calg_3^{}] =
X_{\beta,\gamma}[\calg_3']\cdot X_{\alpha,\gamma}[\calg_2']\cdot
X_{\alpha,\beta}[\calg_1']\;,
\end{equation}
relating the set of 12 variables $\calg_\vrt^{}$ with the set of
12 variables $\calg_\vrt'$, $\vrt=1,2,3$. Equation (\ref{KE})
describes a single 3d vertex. Equations of motion for three
dimensional integrable model is the collection of equations
(\ref{KE}) for all vertices of the 3d lattice.

Korepanov equation needs a very important comment. Matrices
$X_{\#,\#}[\calg_\vrt]$ by definition (\ref{imb}) act in the
direct sum of one-dimensional vector spaces labelled by the
indices $\alpha$, $\beta$, $\gamma$, etc. The matrix
$X_{\alpha,\beta}[\calg_1]$ in the block $(\alpha,\beta)$
coincides with $X[\calg_1]$ (\ref{X-kor}), and in the block
$(\gamma,\dots)$ it is the unity matrix. In what follows, such
``direct sum'' imbedding of $2\times 2$ matrices $X$ into higher
dimensional unity matrices will always be implied.

\subsection{Linear problem with periodical boundary conditions}

\begin{figure}[ht]
\begin{center}
\setlength{\unitlength}{0.25mm}
\begin{picture}(620,250)
\put(0,30){\begin{picture}(620,220)
\thinlines
 \drawline[-30](0,0)(490,210)
 \put(-15,-10){\scriptsize $\alpha_3$}
 \drawline[-30](200,0)(550,210)
 \put(185,-10){\scriptsize $\alpha_2$}
 \drawline[-30](400,0)(610,210)
 \put(390,-10){\scriptsize $\alpha_1$}
 \drawline[-30](25,30)(475,30)
 \put(5,28){\scriptsize $\beta_3$}
 \drawline[-30](212.5,105)(537.5,105)
 \put(191,103){\scriptsize $\beta_2$}
 \drawline[-30](362.6,165)(587.5,165)
 \put(340,163){\scriptsize $\beta_1$}
\Thicklines
\path(70,3)(70,57)\path(245,85.5)(245,124.5)\path(385,151.5)(385,178.5)
\path(250,3)(250,57)\path(375,85.5)(375,124.5)\path(475,151.5)(475,178.5)
\path(430,3)(430,57)\path(505,85.5)(505,124.5)\path(565,151.5)(565,178.5)
\end{picture}}\end{picture}
\end{center}
\caption{A fragment of the auxiliary square lattice.}
\label{fig-TM}
\end{figure}
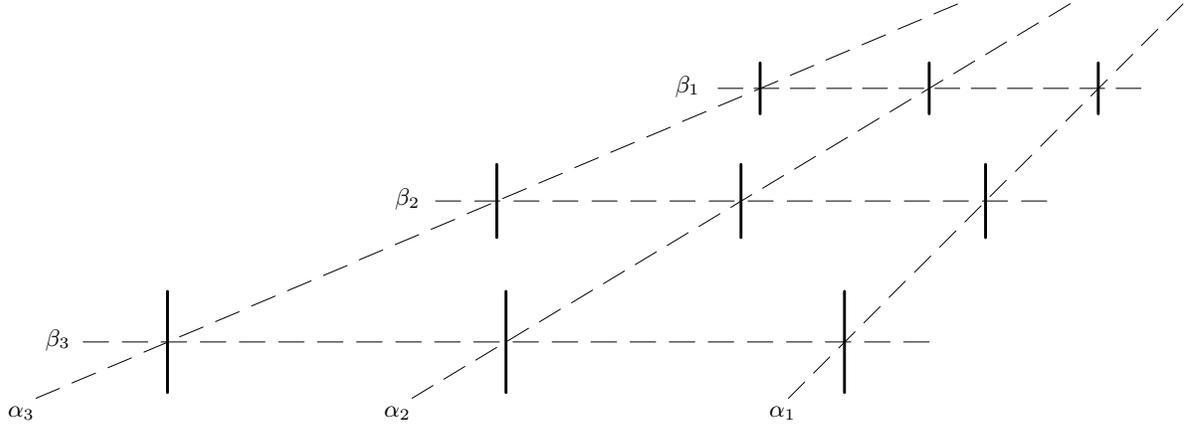

Korepanov's solution of the equations of motion is based on the
solution of the linear problem for the whole auxiliary lattice.
Consider the square lattice with the sizes $N\times M$. Let the
lines of the lattice are enumerated by
\begin{equation}
\alpha_n\quad \textrm{and}\quad \beta_m\;,\quad n=1,2,\dots,
N,\quad m=1,2,\dots, M\;.
\end{equation}
A fragment of the auxiliary lattice is shown in Fig. \ref{fig-TM}.
Notations for vertex and auxiliary variables for
$(n,m)^{\,\textrm{th}}$ vertex of the plane are shown in Fig.
\ref{fig-LP2}. The local auxiliary linear problem (\ref{X-kor})
for $(n,m)^{\,\textrm{th}}$ vertex takes the form
\begin{equation}\label{LP2}
\tvec{\psi_{\alpha_n}^{(m-1)}}{\psi_{\beta_m}^{(n-1)}} =
X[\calg_{n,m}] \;\cdot
\;\tvec{\psi_{\alpha_n}^{(m)}}{\psi_{\beta_m}^{(n)}}\;,\quad
n=1,\dots N,\ m=1,\dots, M\;.
\end{equation}
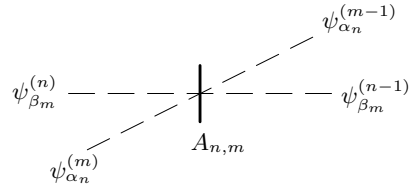
\begin{figure}[ht]
\begin{center}
\setlength{\unitlength}{0.25mm}
\begin{picture}(200,200)
 \thinlines
 \drawline[-30](30,100)(170,100)
 \drawline[-30](40,70)(160,130)
 \Thicklines
 \path(100,85)(100,115)
 \put(0,97){\scriptsize $\psi_{\beta_m}^{(n)}$}
 \put(20,55){\scriptsize $\psi_{\alpha_n}^{(m)}$}
 \put(165,135){\scriptsize $\psi_{\alpha_n}^{(m-1)}$}
 \put(175,95){\scriptsize $\psi_{\beta_m}^{(n-1)}$}
 \put(95,70){\scriptsize $\calg_{n,m}$}
\end{picture}
\end{center}
\caption{Notations for $(n,m)^{\,\textrm{th}}$ vertex of the
auxiliary lattice.} \label{fig-LP2}
\end{figure}
The linearity of the whole set of (\ref{LP2}) with respect to
$\psi$-s makes it possibly to define the quasi-periodical boundary
conditions for them:
\begin{equation}\label{uv-bc}
\psi_{\alpha_n}^{(m+M)} = u \psi_{\alpha_n}^{(m)} \quad
\textrm{and} \quad \psi_{\beta_m}^{(n+N)} = v
\psi_{\beta_m}^{(n)}\;,
\end{equation}
where $u$ and $v$ are $\mathbb{C}$-valued spectral parameters.

Linear equations (\ref{LP2}) may be iterated for the whole lattice
as follows. Let
\begin{equation}\label{Psi}
\psi_{\boldsymbol{\alpha}}^{(m)}\;=\;\left(\begin{array}{c} \psi_{\alpha_1}^{(m)}\\
\psi_{\alpha_2}^{(m)}\\ \vdots\\ \psi_{\alpha_N}^{(m)}
\end{array}\right)\;,
\;\;\;\; \psi_{\boldsymbol{\beta}}^{(n)}\;=\;\left(\begin{array}{c} \psi_{\beta_1}^{(n)}\\
\psi_{\beta_2}^{(n)}\\ \vdots\\
\psi_{\beta_M}^{(n)}\end{array}\right)\;.
\end{equation}
Then the repeated use of (\ref{LP2}) gives
\begin{equation}\label{Psi-X-Psi}
\tvec{\psi_{\boldsymbol{\alpha}}^{(0)}}{\psi_{\boldsymbol{\beta}}^{(0)}}\;=\;
\mathbb{X}_{\boldsymbol{\alpha},\boldsymbol{\beta}}\;
\tvec{\psi_{\boldsymbol{\alpha}}^{(M)}}{\psi_{\boldsymbol{\beta}}^{(N)}}\;,
\end{equation}
where, in the terms of matrix imbedding discussed right after
(\ref{KE}), the $(N+M)\times (N+M)$ monodromy matrix
$\mathbb{X}_{\boldsymbol{\alpha},\boldsymbol{\beta}}$ may be
written as
\begin{equation}\label{big-X}
\mathbb{X}_{\boldsymbol{\alpha},\boldsymbol{\beta}}\;=\;
\prod_{n}^{\curvearrowright}\;\prod_{m}^{\curvearrowright}\;X_{\alpha_n,\beta_m}[\calg_{n,m}]\;,
\end{equation}
where
\begin{equation}
\prod_n^{\curvearrowright}\;f_n\;\stackrel{\textrm{def}}{=}\;
f_1\;f_{2}\;\dots\;f_{N-1}\;f_{N}\;,\;\;\;
\prod_m^{\curvearrowright}\;f_m\;\stackrel{\textrm{def}}{=}\;
f_1\;f_{2}\;\dots\;f_{M-1}\;f_{M}\;.
\end{equation}
The boundary conditions (\ref{uv-bc}) give
$\psi_{\boldsymbol{\alpha}}^{(M)}=u\psi_{\boldsymbol{\alpha}}^{(0)}$,
$\psi_{\boldsymbol{\beta}}^{(N)}=v\psi_{\boldsymbol{\beta}}^{(0)}$,
so that (\ref{Psi-X-Psi}) becomes
\begin{equation}
\left(1\;-\;\mathbb{X}_{\boldsymbol{\alpha},\boldsymbol{\beta}} \;
\cdot\;\left(\begin{array}{cc} u & 0\\ 0 & v
\end{array}\right)\right)
\tvec{\psi_{\boldsymbol{\alpha}}^{(0)}}{\psi_{\boldsymbol{\beta}}^{(0)}}\;=\;0\;.
\end{equation}
The whole linear problem has a solution if and only if
\begin{equation}\label{determinant}
J(u,v)\;\stackrel{\textrm{def}}{=}\;
\det\left(1\;-\;\mathbb{X}_{\boldsymbol{\alpha},\boldsymbol{\beta}}
\;\cdot\;\left(\begin{array}{cc} u & 0\\
0 & v
\end{array}\right) \right)
\end{equation}
is zero. Equation $J(u,v)=0$ defines the spectral curve for the
model, equations of motion (\ref{KE}) for the whole three
dimensional lattice have an exact solution in the terms of
$\theta$-functions on the Jacobian of the spectral curve
\cite{Korepanov}.

\subsection{Free fermion model.}

The determinant (\ref{determinant}) has the very well known
combinatorial representation. Usual way to derive it is to define
the determinant in the terms of the Grassmanian integration and
then to turn from normal symbols to matrix elements.

Let in this subsection $\psi$ and $\overline{\psi}$ be the
Grassmanian variables with the integration rules $\ds\int
d\psi=\int d\overline{\psi}=0$ and $\ds\int\psi
d\psi=\int\overline{\psi}d\overline{\psi}=1$. Then the determinant
(\ref{determinant}) may be written as
\begin{equation}\label{39}
\ds
J(u,v)\;=\;\int\;\EXP^{\mathscr{A}[\overline{\psi},\psi]}\;\mathscr{D}\overline{\psi}\mathscr{D}\psi\;,
\end{equation}
where the ``action'' is
\begin{equation}\label{action}
\mathscr{A}\;=\;\sum_{n,m=1}^{N,M}\left\{
\left(\overline{\psi}_{\alpha_n}^{(m-1)},\overline{\psi}_{\beta_m}^{(n-1)}\right)
\cdot X[\calg_{n,m}] \cdot
\left(\begin{array}{cc}\psi_{\alpha_n}^{(m)}\\
\psi_{\beta_m}^{(n)}\end{array}\right) \;+\;
\psi_{\alpha_n}^{(m)}\overline{\psi}_{\alpha_n}^{(m)} \;+\;
\psi_{\beta_m}^{(n)}\overline{\psi}_{\beta_m}^{(n)}\right\}\;,
\end{equation}
and the measure is
\begin{equation}
\mathscr{D}\overline{\psi}\mathscr{D}\psi\;=\; \prod_{n,m=1}^{N,M}
d\overline\psi_{\alpha_n}^{(m)}d\psi_{\alpha_n}^{(m)}
d\overline\psi_{\beta_m}^{(n)} d\psi_{\beta_m}^{(n)}\;.
\end{equation}
Spectral parameters appear in (\ref{39})  via
\begin{equation}
\overline{\psi}_{\alpha_n}^{(0)}\;=\;u\overline{\psi}_{\alpha_n}^{(M)}
\quad \textrm{and} \quad
\overline{\psi}_{\beta_m}^{(0)}\;=\;v\overline{\psi}_{\beta_m}^{(N)}\;.
\end{equation}

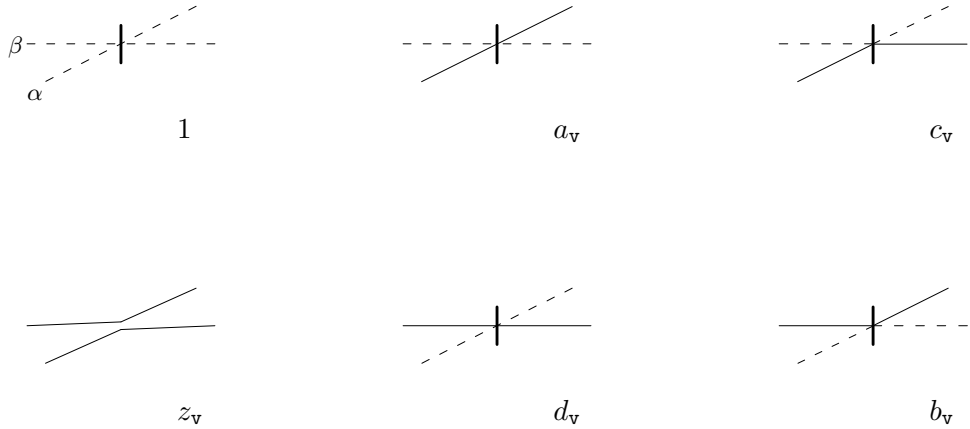
\begin{figure}[ht]
\begin{center}
\setlength{\unitlength}{0.25mm}
\begin{picture}(500,300)
\thinlines
\put(0,200){\begin{picture}(100,100)
 \thinlines
 \dashline{5}(0,50)(100,50)
 \dashline{5}(10,30)(90,70)
 \Thicklines\path(50,40)(50,60)
 \put(80,0){\small $1$}
 \put(-10,45){\scriptsize $\beta$}
 \put(0,20){\scriptsize $\alpha$}
\end{picture}}
\put(0,50){\begin{picture}(100,100)
 \thinlines
 \drawline(0,50)(50,52)(90,70)
 \drawline(10,30)(50,48)(100,50)
 \put(80,0){\small $z_\vrt$}
\end{picture}}
\put(200,200){\begin{picture}(100,100)
 \thinlines
 \dashline{5}(0,50)(100,50)
 \drawline(10,30)(90,70)
 \Thicklines\path(50,40)(50,60)
 \put(80,0){\small $a_\vrt$}
\end{picture}}
\put(200,50){\begin{picture}(100,100)
 \thinlines
 \drawline(0,50)(100,50)
 \dashline{5}(10,30)(90,70)
 \Thicklines\path(50,40)(50,60)
 \put(80,0){\small $d_\vrt$}
\end{picture}}
\put(400,200){\begin{picture}(100,100)
 \thinlines
 \dashline{5}(0,50)(50,50)(90,70)
 \drawline(10,30)(50,50)(100,50)
 \Thicklines\path(50,40)(50,60)
 \put(80,0){\small $c_\vrt$}
\end{picture}}
\put(400,50){\begin{picture}(100,100)
 \thinlines
 \drawline(0,50)(50,50)(90,70)
 \dashline{5}(10,30)(50,50)(100,50)
 \Thicklines\path(50,40)(50,60)
 \put(80,0){\small $b_\vrt$}
\end{picture}}
\end{picture}
\end{center}
\caption{Six variants of bypassing the vertex. Vertex factors
$f_\vrt$ are matrix elements of $L$. Note, in the variant $z_\vrt$
the path is not self-intersecting.} \label{fig-fragments}
\end{figure}

In the terms of Grassmanian variables, the exponent of a quadratic
form is a normal symbol of some operator $L$,
\begin{equation}\label{Ldef}
\exp\left\{
\left(\overline{\psi}_{\alpha},\overline{\psi}_{\beta}\right)
\cdot X[\calg_\vrt]\cdot
\left(\begin{array}{cc}\psi_{\alpha}\\
\psi_{\beta}\end{array}\right)\right\}
\;\stackrel{\textrm{def}}{=}\; \langle
\overline{\psi}_\alpha,\overline{\psi}_\beta|
L_{\alpha,\beta}[\calg_\vrt] |\psi_\alpha,\psi_\beta\rangle\;.
\end{equation}
The fermionic coherent states are defined by
\begin{equation}\label{44}
|\psi\rangle\;=\;|0\rangle +|1\rangle \psi\;,\;\;\;
\langle\overline{\psi}|\;=\;\langle 0|+\overline{\psi}\langle
1|\;,
\end{equation}
and the extra summands in (\ref{action}) correspond to the unity
operators
\begin{equation}
1\;=\;\int |\psi\rangle \;\EXP^{\psi\overline{\psi}}
\;d\overline{\psi}\;d\psi \;\langle\overline{\psi}|\;.
\end{equation}
It is important to note that the indices of the operator
$L_{\alpha,\beta}[\calg_\vrt]$ (\ref{Ldef}) label copies of
two-dimensional vector spaces (\ref{44}) $\mathbb{C}^2\ni
x|0\rangle+y|1\rangle$ . Thus, $L_{\alpha,\beta}$ acts in the
tensor \emph{product} of two-dimensional vector spaces, while
$X_{\alpha,\beta}$ acts in the tensor \emph{sum} of
one-dimensional vector spaces. In the basis of the fermionic
states
\begin{equation}\label{ordering}
|n_\alpha,n_\beta\rangle \;=\; \biggl(|0,0\rangle, |1,0\rangle,
|0,1\rangle, |1,1\rangle\biggr)\;,
\end{equation}
operator $L_{\alpha,\beta}$ (\ref{Ldef}) is $4\times 4$ matrix
\begin{equation}\label{L-kor}
L_{\alpha,\beta}[\calg_\vrt]\;=\;\left(\begin{array}{cccc} 1 & 0 & 0 & 0 \\
0 & a_\vrt & b_\vrt & 0\\
0 & c_\vrt & d_\vrt & 0\\
0 & 0 & 0 & z_\vrt\end{array}\right)\;,\;\;\; \textrm{where}\;\;\;
z_\vrt\;\stackrel{\textrm{def}}{=}\;b_\vrt c_\vrt - a_\vrt
d_\vrt\;.
\end{equation}
Besides, the Korepanov equation (\ref{KE}) is the equality of the
exponents of the normal symbol form of the local Yang-Baxter
equation
\begin{equation}\label{lybe}
L_{\alpha,\beta}[\calg_1^{}] L_{\alpha,\gamma}[\calg_2^{}]
L_{\beta,\gamma}[\calg_3^{}] \;=\; L_{\beta,\gamma}[\calg_3']
L_{\alpha,\gamma}[\calg_2'] L_{\alpha,\beta}[\calg_3']\;,
\end{equation}
since
\begin{equation}\label{xxx-lll}
\langle\overline{\psi}|L_{\alpha,\beta}L_{\alpha,\gamma}L_{\beta,\gamma}|\psi\rangle
\;=\; \exp\left\{\overline{\psi} \cdot
X_{\alpha,\beta}X_{\alpha,\gamma}X_{\beta,\gamma} \cdot
\psi\right\}\;\;\;\;\textrm{etc.}
\end{equation}
Turn now to the expression of the determinant (\ref{determinant})
in the terms of operators $L$. Let $2^{N+M}\times 2^{N+M}$ matrix
$\mathbb{L}_{\boldsymbol{\alpha},\boldsymbol{\beta}}$ be the
ordered product of local $L$-s:
\begin{equation}\label{big-L}
\mathbb{L}_{\boldsymbol{\alpha},\boldsymbol{\beta}}\;=\;
\prod_n^{\curvearrowright}\prod_m^{\curvearrowright}\;L_{\alpha_n,\beta_m}[\calg_{n,m}]\;.
\end{equation}
This is related to the monodromy matrix (\ref{big-X}) by means of
(cf. (\ref{Ldef}))
\begin{equation}
\exp\left\{
\left(\overline{\psi}_{\boldsymbol{\alpha}},\overline{\psi}_{\boldsymbol{\beta}}\right)
\cdot \mathbb{X}_{\boldsymbol{\alpha},\boldsymbol{\beta}}\cdot
\left(\begin{array}{cc}\psi_{\boldsymbol{\alpha}}\\
\psi_{\boldsymbol{\beta}}\end{array}\right)\right\} \;=\;
\langle\overline{\psi}_{\boldsymbol{\alpha}},\overline{\psi}_{\boldsymbol{\beta}}|
\mathbb{L}_{\boldsymbol{\alpha},\boldsymbol{\beta}}|\psi_{\boldsymbol{\alpha}},\psi_{\boldsymbol{\beta}}\rangle\;.
\end{equation}
Define now the boundary matrices for
$\mathbb{L}_{\boldsymbol{\alpha},\boldsymbol{\beta}}$,
\begin{equation}\label{52}
D(u)\;\stackrel{\textrm{def}}{=}\;\left(\begin{array}{cc} 1 & 0 \\
0 & u
\end{array}\right)\;,\;\;\;\;
D_{\boldsymbol{\alpha}}(u)\;=\;\prod_{n} D_{\alpha_n}(u)\;,\;\;\;
D_{\boldsymbol{\beta}}(v)\;=\;\prod_{m} D_{\beta_m}(v)\;,
\end{equation}
and let
\begin{equation}\label{T-matrix}
T(u,v)\;=\;\mathop{\textrm{Trace}}_{\boldsymbol{\alpha},\boldsymbol{\beta}}
\biggl(D_{\boldsymbol{\alpha}}(u)D_{\boldsymbol{\beta}}(v)
\mathbb{L}_{\boldsymbol{\alpha},\boldsymbol{\beta}}\biggr)\;.
\end{equation}
By the construction, $T(u,v)$ is the partition function for a
free-fermion lattice model with the inhomogeneous Boltzmann
weights -- matrix elements of $L_{\alpha_n,\beta_m}[\calg_{n,m}]$
-- and $u,v$-boundary conditions. It is the polynomial of $u$ and
$v$:
\begin{equation}\label{T-decomp}
T(u,v)\;=\;\sum_{n=0}^{N} \sum_{m=0}^{M} u^n v^m t_{n,m}\;.
\end{equation}
Sometimes a pure combinatorial representation of the partition
function is very useful. Any monomial in $T(u,v)$ (\ref{T-matrix})
corresponds to a non-self-intersecting path on the toroidal
lattice. A path may go through a vertex in one of five different
ways as it is shown in Fig. \ref{fig-fragments} (or do not go
through at all). A factor $f_\vrt$ is associated with each
variant, these factors are the matrix elements of
$L_{\alpha,\beta}[\calg_\vrt]$ (\ref{L-kor}).
A monomial in $T(u,v)$, corresponding to path $C$, is
\begin{equation}\label{t_C}
t_{C}=\prod_{\textrm{along path}\; C} f_\vrt\;.
\end{equation}
Any non-self-intersecting path on the toroidal lattice has a
homotopy class
\begin{equation}
w(C)=n A + m B\;,
\end{equation}
where $A$ is the toroidal cycle along the $\alpha$-lines, and $B$
is the toroidal cycle along the $\beta$-lines of Fig.
\ref{fig-TM}. Then the element $t_{n,m}$ of (\ref{T-decomp}) is
\begin{equation}\label{t_nm}
t_{n,m}\;=\;\sum_{C\;:\;w(C)=nA+mB} t_{C}
\end{equation}
The determinant $J(u,v)$ (\ref{determinant}) is related to
$t_{n,m}$ via
\begin{equation}\label{J-T-classic}
J(u,v)\;=\;\sum_{n=0}^{N} \sum_{m=0}^{M} (-)^{nm+n+m} u^n v^m
t_{n,m}\;,
\end{equation}
where the sign $(-)^{nm+n+m}$ counts the number of fermionic loops
on the toroidal square lattice. The determinant $J$ may be
expressed in the terms of $T$ and vice versa:
\begin{equation}\label{56}
\begin{array}{l} \ds J(u,v)\;=\;\frac{1}{2}\left(
T(-u,-v)+T(-u,v)+T(u,-v)-T(u,v)\right)\;,\\
\\
\ds T(u,v)\;=\;\frac{1}{2}\left(
J(-u,-v)+J(-u,v)+J(u,-v)-J(u,v)\right)\;.
\end{array}
\end{equation}
The last equality is very well known in the two dimensional
free-fermion model as the formula relating the lattice partition
function and fermionic determinant.

Now we may finish the collection of notions and definitions of the
Korepanov model of classical integrable dynamics on three
dimensional lattice and proceed to the description of their
quantum analogues.

\section{Quantum model}

In the previous section, we did not pay any attention to the
structure of vertex variables $\calg_\vrt$ (\ref{abcd}), they were
defined simply as the list of elements of $X$ (\ref{X-kor}). The
key point for the quantization of the model is that it is possible
do define a local Poisson structure on $\calg_\vrt$ \cite{First}
such that the transformation
\begin{equation}
\calg_1^{}\otimes \calg_2^{}\otimes \calg_3^{} \; \to \;
\calg_1'\otimes \calg_2'\otimes \calg_3'\;,
\end{equation}
defined by Korepanov equation (\ref{KE}), is a symplectic map.
Symplectic structure admits an immediate quantization. We will
skip here all the details and proceed directly to the Ansatz for
quantized $\calg_\vrt$, $X[\calg_\vrt]$ and $L[\calg_\vrt]$. The
aim of this section is just to give precise definition of
$\mathbf{T}(u,v)$ (\ref{transfer-matrix}).

\subsection{Quantum Korepanov and Tetrahedron equations}

The local $q$-oscillator algebra $\alg$ is defined by
(\ref{q-osc}). The Fock space $\mathcal{F}$ representation for
$q$-oscillators corresponds to
\begin{equation}\label{fock}
\textrm{Spec}( \Nop_\vrt ) \;=\; 0,1,2,\dots \quad \forall\
\vrt\;.
\end{equation}
Quantized dynamical variables (\ref{abcd}) are the $q$-oscillator
generators and a pair of $\mathbb{C}$-valued parameters,
$\qalg_\vrt\sim (\alg_\vrt; \lambda_\vrt,\mu_\vrt)$:
\begin{equation}\label{q-abcd}
\qalg_\vrt \;=\; \left(a_\vrt=\lambda_\vrt
q^{\Nop_\vrt},\;b_\vrt=\yop_\vrt,\;
c_\vrt=-q^{-1}\lambda_\vrt\mu_\vrt\xop_\vrt,\; d_\vrt=\mu_\vrt
q^{\Nop_\vrt}\right)\;.
\end{equation}
Quantized $X$ (\ref{X-kor}) and $L$
(\ref{L-kor}) are given by
\begin{equation}\label{quantum-X}
X[\qalg_\vrt]\;=\;\left(\begin{array}{cc} \lambda_\vrt q^{\Nop_\vrt} & \yop_\vrt\\
-q^{-1}\lambda_\vrt\mu_\vrt\xop_\vrt & \mu_\vrt
q^{\Nop_\vrt}\end{array}\right)
\end{equation}
and
\begin{equation}\label{quantum-L}
L_{\alpha,\beta}[\qalg_\vrt]\;=\;\left(\begin{array}{cccc} 1 & 0 & 0 & 0 \\
0 & \lambda_\vrt q^{\Nop_\vrt} & \yop_\vrt & 0 \\
0 & -q^{-1}\lambda_\vrt\mu_\vrt\xop_\vrt & \mu_\vrt q^{\Nop_\vrt} & 0 \\
0 & 0 & 0 & -q^{-1}\lambda_\vrt\mu_\vrt
\end{array}\right)\;.
\end{equation}
One may verify directly, the quantum Korepanov equation
\begin{equation}\label{qke}
X_{\alpha,\beta}[\qalg_1] X_{\alpha,\gamma}[\qalg_2]
X_{\beta,\gamma}[\qalg_3] \ \Rop \;=\; \Rop \
X_{\beta,\gamma}[\qalg_3] X_{\alpha,\gamma}[\qalg_2]
X_{\alpha,\beta}[\qalg_1]
\end{equation}
is equivalent to the auxiliary tetrahedron equation (the quantum
local Yang-Baxter equation)
\begin{equation}\label{qlybe}
L_{\alpha,\beta}[\qalg_1] L_{\alpha,\gamma}[\qalg_2]
L_{\beta,\gamma}[\qalg_3] \ \Rop \;=\; \Rop \
L_{\beta,\gamma}[\qalg_3] L_{\alpha,\gamma}[\qalg_2]
L_{\alpha,\beta}[\qalg_1]\;.
\end{equation}
Here the intertwining operator $\Rop=\Rop_{123}$ acts in the
tensor product of representation spaces $\mathcal{F}_1 \otimes
\mathcal{F}_2 \otimes \mathcal{F}_3$ of three $q$-oscillators
$\alg_{1,2,3}$. Parameters $\lambda_\vrt,\mu_\vrt$ of $\qalg_\vrt$
are the parameters of $\Rop$. Both equations (\ref{qke}) and
(\ref{qlybe}) are equivalent to the following set of six
equations:
\begin{equation}\label{mapping}
\begin{array}{ll}
\ds \Rop \; q^{\Nop_2}\xop_1^{} =
\frac{\lambda_2}{\lambda_3}\biggl(
q^{\Nop_3}\xop_1^{}-\frac{q}{\lambda_1\mu_3}
q^{\Nop_1}\xop_2^{}\yop_3\biggr)\;\Rop\;,&
\ds\Rop\;\xop_2^{} =
\biggl(\xop_1^{}\xop_3^{}\,+\,\frac{q^2}{\lambda_1\mu_3}
q^{\Nop_1+\Nop_3}\xop_2^{}\biggr) \; \Rop\;, \\
\ds \Rop \; q^{\Nop_2}\xop_3^{} = \frac{\mu_2}{\mu_1} \biggl(
q^{\Nop_1}\xop_3^{} - \frac{q}{\lambda_1\mu_3}
q^{\Nop_3}\yop_1^{}\xop_2^{}\biggr) \; \Rop\;,&
\ds \Rop\; \yop_2^{} =
\biggl(\yop_1^{}\yop_3^{}\,+\,\lambda_1^{}\mu_3^{}
q^{\Nop_1+\Nop_3}\yop_2^{}\biggr) \; \Rop\;, \\
\ds \Rop q^{\Nop_1+\Nop_2} \;=\; q^{\Nop_1+\Nop_2} \Rop\;,& \Rop
q^{\Nop_2+\Nop_3}\;=\; q^{\Nop_2+\Nop_3} \Rop\;.
\end{array}
\end{equation}
Classical equations (\ref{KE}) and (\ref{lybe}) follow from
(\ref{qke}) and (\ref{qlybe}) in the $q\to 1$ limit of the well
defined automorphism $\qalg_\vrt'=\Rop\qalg_\vrt^{}\Rop^{-1}$. For
irreducible representations of $\alg_\vrt$, $\Rop$ is defined
uniquely, its matrix elements for the Fock space representation
are given in \cite{First}. Remarkably, Fig. \ref{fig-palms} may be
used for the graphical representation both of classical and
quantum equations, in the quantum case the solid $3d$ cross in
Fig. \ref{fig-palms} stands for $\Rop$.

Integrable model of quantum mechanics may be formulated purely in
the terms of matrices $L$ (\ref{quantum-L}). The intertwiner
$\Rop$ is related to evolution operators, it is another subject
and we will not consider it here. Below we recall the definition
of integrable model of quantum mechanics from \cite{First,letter}.

\subsection{Transfer matrix $\mathbf{T}$}

For the square lattice $N\times M$ of the previous section, define
the ``monodromy'' of quantized $L$-s literally by (\ref{big-L}):
\begin{equation}\label{q-big-L}
\mathbb{L}_{\boldsymbol{\alpha},\boldsymbol{\beta}}\;=\;
\prod_n^{\curvearrowleft}\prod_m^{\curvearrowleft}\;L_{\alpha_n,\beta_m}[\qalg_{n,m}]\;,
\end{equation}
and its trace (cf. (\ref{T-matrix}))
\begin{equation}\label{q-T-matrix}
\mathbf{T}(u,v)\;=\;\mathop{\textrm{Trace}}_{\boldsymbol{\alpha},\boldsymbol{\beta}}
\biggl(D_{\boldsymbol{\alpha}}(u)D_{\boldsymbol{\beta}}(v)
\mathbb{L}_{\boldsymbol{\alpha},\boldsymbol{\beta}}\biggr)\;,
\end{equation}
where boundary matrices $D$ are defined by (\ref{52}). To
distinguish the classical and quantum cases, we use the boldface
letters for the quantized $\mathbf{T}$ and its decomposition
(\ref{T-decomp}):
\begin{equation}\label{q-T-decomp}
\mathbf{T}(u,v)\;=\;\sum_{n=0}^{N} \sum_{m=0}^{M} u^n v^m
\mathbf{t}_{n,m}\;.
\end{equation}
Since the arguments of $L$-s are local $q$-oscillator generators,
$\mathbf{T}(u,v)\in\alg^{\otimes NM}$ is by definition a
layer-to-layer transfer matrix, its graphical representation is
again the classical Fig. \ref{fig-TM}. The model is integrable
since
\begin{equation}\label{TT-TT}
\mathbf{T}(u,v)\;\mathbf{T}(u',v')\;=\;\mathbf{T}(u',v')\;\mathbf{T}(u,v)\;,
\end{equation}
i.e. the coefficients $\mathbf{t}_{n,m}$ in (\ref{q-T-decomp})
form the set of the integrals of motion.

Now we give the technical proof of the commutativity
(\ref{TT-TT}). The commutativity of layer-to-layer transfer
matrices follow from a proper tetrahedron equation \cite{bs}. In
addition to $L_{\alpha,\beta}[\qalg_\vrt]$ (\ref{quantum-L}),
define
\begin{equation}
\widetilde{L}_{\alpha,\beta}[\qalg_0]\;=\;\left(\begin{array}{cccc} 1 & 0 & 0 & 0 \\
0 & \lambda_0 (-q)^{\Nop_0} & \yop_0 & 0 \\
0 & q^{-1}\lambda_0\mu_0\xop_0 & \mu_0 (-q)^{\Nop_0} & 0 \\
0 & 0 & 0 & q^{-1}\lambda_0\mu_0
\end{array}\right)\;,
\end{equation}
where $\qalg_0\sim (\alg_0;\lambda_0,\mu_0)$. The \emph{constant}
tetrahedron equation for $L$ and $\widetilde{L}$,
\begin{equation}\label{LTE}
\ds \widetilde{L}_{\alpha,\alpha'}[\qalg_0]
\widetilde{L}_{\beta,\beta'}[\qalg_0] L_{\alpha,\beta}[\qalg]
L_{\alpha',\beta'}[\qalg]=L_{\alpha',\beta'}[\qalg]
L_{\alpha,\alpha'}[\alg] \widetilde{L}_{\beta,\beta'}[\qalg_0]
\widetilde{L}_{\alpha,\alpha'}[\qalg_0]
\end{equation}
may be verified directly, it is just $16\times 16$ matrix equation
with the operator-valued entries. Another technical relation is
\begin{equation}\label{DDL}
D_\alpha(u)D_{\alpha'}(u')
\widetilde{L}_{\alpha,\alpha'}[\qalg_0]\;=\;
\left(\frac{u}{u'}\right)^{\Nop_0}
\widetilde{L}_{\alpha,\alpha'}[\qalg_0]
\left(\frac{u'}{u}\right)^{\Nop_0} D_\alpha(u) D_{\alpha'}(u')\;,
\end{equation}
where $D$ is given by (\ref{52}). Combining (\ref{LTE}) for the
whole lattice, we come to
\begin{equation}\label{LTE2}
\widetilde{\mathbb{L}}_{\boldsymbol{\alpha},\boldsymbol{\alpha}'}\;
\widetilde{\mathbb{L}}_{\boldsymbol{\beta},\boldsymbol{\beta}'}\;
\mathbb{L}_{\boldsymbol{\alpha},\boldsymbol{\beta}}\;
\mathbb{L}_{\boldsymbol{\alpha}',\boldsymbol{\beta}'} \;=\;
\mathbb{L}_{\boldsymbol{\alpha}',\boldsymbol{\beta}'}\;
\mathbb{L}_{\boldsymbol{\alpha},\boldsymbol{\beta}}\;
\widetilde{\mathbb{L}}_{\boldsymbol{\beta},\boldsymbol{\beta}'}\;
\widetilde{\mathbb{L}}_{\boldsymbol{\alpha},\boldsymbol{\alpha}'}\;,
\end{equation}
where besides the ``monodromies'' (\ref{big-L}) of $L[\calg_\vrt]$
\begin{equation}
\mathbb{L}_{\boldsymbol{\alpha},\boldsymbol{\beta}}\;=\;
\prod_{n,m}^{\curvearrowright}
L_{\alpha_n,\beta_m}[\qalg_{n,m}]\;\;\textrm{and}\;\;
\mathbb{L}_{\boldsymbol{\alpha}',\boldsymbol{\beta}'}\;=\;
\prod_{n,m}^{\curvearrowright}
L_{\alpha_n',\beta_m'}[\qalg_{n,m}]\;,
\end{equation}
we used
\begin{equation}
\widetilde{\mathbb{L}}_{\boldsymbol{\alpha},\boldsymbol{\alpha}'}\;=\;
\prod_n^\curvearrowright
\widetilde{L}_{\alpha_n,\alpha_n'}[\qalg_0]\;\;\textrm{and}\;\;
\widetilde{\mathbb{L}}_{\boldsymbol{\beta},\boldsymbol{\beta}'}\;=\;
\prod_m^\curvearrowright
\widetilde{L}_{\beta_m,\beta_m'}[\qalg_0]\;.
\end{equation}
Multiplying (\ref{LTE2}) by $D_{\boldsymbol{\alpha}}(u)
D_{\boldsymbol{\beta}}(v) D_{\boldsymbol{\alpha}'}(u')
D_{\boldsymbol{\beta}'}(v')$ and taking into account (\ref{DDL}),
we get
\begin{equation}
\begin{array}{l}
\ds
\widetilde{\mathbb{L}}_{\boldsymbol{\alpha},\boldsymbol{\alpha}'}
\left(\frac{u'v}{uv'}\right)^{\Nop_0}
\widetilde{\mathbb{L}}_{\boldsymbol{\beta},\boldsymbol{\beta}'}\;\cdot\;
D_{\boldsymbol{\alpha}}(u)D_{\boldsymbol{\beta}}(v)
\mathbb{L}_{\boldsymbol{\alpha},\boldsymbol{\beta}}\;\cdot\;
D_{\boldsymbol{\alpha'}}(u') D_{\boldsymbol{\beta}'}(v')
\mathbb{L}_{\boldsymbol{\alpha}',\boldsymbol{\beta}'} \;=\\
\\
\ds  D_{\boldsymbol{\alpha'}}(u') D_{\boldsymbol{\beta}'}(v')
\mathbb{L}_{\boldsymbol{\alpha}',\boldsymbol{\beta}'}\;\cdot\;
D_{\boldsymbol{\alpha}}(u)D_{\boldsymbol{\beta}}(v)
\mathbb{L}_{\boldsymbol{\alpha},\boldsymbol{\beta}}\;\cdot\;
\left(\frac{u'}{u}\right)^{\Nop_0}
\widetilde{\mathbb{L}}_{\boldsymbol{\beta},\boldsymbol{\beta}'}
\widetilde{\mathbb{L}}_{\boldsymbol{\alpha},\boldsymbol{\alpha}'}
\left(\frac{v}{v'}\right)^{\Nop_0}\;.
\end{array}
\end{equation}
Now taking the trace over the representation space $\mathcal{F}_0$
of $\alg_0$ and denoting
\begin{equation}
\mathbb{M}_{\boldsymbol{\alpha},\boldsymbol{\beta},\boldsymbol{\alpha}',\boldsymbol{\beta}'}\;=\;
\mathop{\textrm{Trace}}_{\mathcal{F}_0}\biggl(
\widetilde{\mathbb{L}}_{\boldsymbol{\alpha},\boldsymbol{\alpha}'}
\left(\frac{u'v}{uv'}\right)^{\Nop_0}
\widetilde{\mathbb{L}}_{\boldsymbol{\beta},\boldsymbol{\beta}'}\biggr)\;,
\end{equation}
we come to the final similarity relation
\begin{equation}\label{MLL}
\begin{array}{l}
\ds
\mathbb{M}_{\boldsymbol{\alpha},\boldsymbol{\beta},\boldsymbol{\alpha}',\boldsymbol{\beta}'}\;\cdot\;
D_{\boldsymbol{\alpha}}(u)D_{\boldsymbol{\beta}}(v)
\mathbb{L}_{\boldsymbol{\alpha},\boldsymbol{\beta}}\;\cdot\;
D_{\boldsymbol{\alpha'}}(u') D_{\boldsymbol{\beta}'}(v')
\mathbb{L}_{\boldsymbol{\alpha}',\boldsymbol{\beta}'} \;=\\
\\
\ds  D_{\boldsymbol{\alpha'}}(u') D_{\boldsymbol{\beta}'}(v')
\mathbb{L}_{\boldsymbol{\alpha}',\boldsymbol{\beta}'}\;\cdot\;
D_{\boldsymbol{\alpha}}(u)D_{\boldsymbol{\beta}}(v)
\mathbb{L}_{\boldsymbol{\alpha},\boldsymbol{\beta}}\;\cdot\;
\mathbb{M}_{\boldsymbol{\alpha},\boldsymbol{\beta},\boldsymbol{\alpha}',\boldsymbol{\beta}'}
\;,
\end{array}
\end{equation}
and therefore two transfer matrices $\ds
\mathbf{T}(u,v)=\mathop{\textrm{Trace}}_{\boldsymbol{\alpha},\boldsymbol{\beta}}
\biggl(D_{\boldsymbol{\alpha}}(u)D_{\boldsymbol{\beta}}(v)
\mathbb{L}_{\boldsymbol{\alpha},\boldsymbol{\beta}}\biggr)$
commute.

In the limit $q\to 1$ coefficients of (\ref{q-T-decomp}) become
the involutive moduli of the spectral curve (\ref{determinant}).
Since the moduli of the classical spectral curve are independent
and the spectral curve defines completely the solution of the
classical model \cite{Korepanov}, the set of integrals of motion
$\textbf{t}_{n,m}$ is complete.

The combinatorial representation for $\mathbf{T}(u,v)$
(\ref{q-T-matrix}) is equivalent to the combinatorial
representation for $T(u,v)$ (\ref{T-matrix}). The vertex factors
of $L[\calg_\vrt]$ (\ref{L-kor}) in Fig. \ref{fig-fragments} are
to be replaced by corresponding elements of $L[\qalg_\vrt]$
(\ref{quantum-L}). Equations (\ref{t_C}) and (\ref{t_nm}) remain
unchanged. In particular, with the help of the combinatorial
representation one may easily see
\begin{equation}\label{T-boundary}
\mathbf{T}(u,0)\;=\;\prod_n\left( 1+u\prod_m\lambda_{n,m}
q^{\Nop_{n,m}}\right)\;,\;\;\; \mathbf{T}(0,v)\;=\;\prod_m\left(
1+v \prod_n\mu_{n,m}q^{\Nop_{n,m}}\right)\;,
\end{equation}
so that the basic integrals of motion are $q^{\mathcal{J}_n}$ and
$q^{\mathcal{K}_m}$,
\begin{equation}\label{charges}
\mathcal{J}_n\;=\;\sum_{m}\;\Nop_{n,m}\;\;\textrm{and}\;\;
\mathcal{K}_m\;=\;\sum_n\;\Nop_{n,m}\;.
\end{equation}
Eigenvalues of $\mathcal{J}_n$ and $\mathcal{K}_m$ fix a subspace
in the state space of the model.

\subsection{Transfer matrix (\ref{q-T-matrix}) and $2d$
quantum inverse scattering method}

The transfer matrix (\ref{q-T-matrix}) may be identically
rewritten as the trace of $2d$ monodromy matrix
\begin{equation}
\mathbf{T}(u,v)\;=\;\mathop{\textrm{Trace}}_{\boldsymbol{\beta}}
\left( D_{\boldsymbol{\beta}}(v)\;\prod_n^\curvearrowright
\mathfrak{L}^{(n)}_{\boldsymbol{\beta}}(u)\right)\;,
\end{equation}
where
\begin{equation}\label{Lax-L}
\mathfrak{L}_{\boldsymbol{\beta}}^{(n)}(u)\;=\;\mathop{\textrm{Trace}}_{\alpha_n}
\biggl(D_{\alpha_n}^{}(u)\prod_m^{\curvearrowright}
L_{\alpha_n,\beta_m}[\qalg_{n,m}]\biggr)\;.
\end{equation}
Equations (\ref{LTE}) and (\ref{DDL}) provide in particular
\begin{equation}
\begin{array}{l}
\ds \widetilde{L}_{\alpha,\alpha'}^{(0)}\;\cdot\;
\left(\frac{u'}{u}\right)^{\Nop_0}\widetilde{L}_{\beta_m,\beta_m'}^{(0)}
\;\cdot\; D_{\alpha}^{}(u)L_{\alpha,\beta_m}^{(m)} \;\cdot\;
D_{\alpha'}^{}(u')L_{\alpha',\beta_m'}^{(m)} \;=\;\\
\\
\ds  D_{\alpha'}^{}(u')L_{\alpha',\beta_m'}^{(m)} \;\cdot\;
D_{\alpha}^{}(u)L_{\alpha,\beta_m}^{(m)} \;\cdot\;
\left(\frac{u'}{u}\right)^{\Nop_0}\widetilde{L}_{\beta_m,\beta_m'}^{(0)}
\;\cdot\; \widetilde{L}_{\alpha,\alpha'}^{(0)}\;,
\end{array}
\end{equation}
and therefore
\begin{equation}
\mathfrak{R}_{\boldsymbol{\beta},\boldsymbol{\beta}'}(u/u')
\mathfrak{L}_{\boldsymbol{\beta}}(u)
\mathfrak{L}_{\boldsymbol{\beta}'}(u')\;=\;
\mathfrak{L}_{\boldsymbol{\beta}'}(u')
\mathfrak{L}_{\boldsymbol{\beta}}(u)
\mathfrak{R}_{\boldsymbol{\beta},\boldsymbol{\beta}'}(u/u')\;,
\end{equation}
where
\begin{equation}\label{big-R}
\mathfrak{R}_{\boldsymbol{\beta},\boldsymbol{\beta}'}(u/u')\;=\;
\mathop{\textrm{Trace}}_{\mathcal{F}_0}\left\{\left(\frac{u'}{u}\right)^{\Nop_0}
\prod_m^{\curvearrowright}
\widetilde{L}_{\beta_m,\beta_m'}[\qalg_0]\right\}\;.
\end{equation}

In the same way, the other direction of the lattice may be chosen,
and the dual Lax operator
\begin{equation}\label{Lax-L2}
\mathfrak{L}_{\boldsymbol{\alpha}}^{(m)}(v)\;=\;\mathop{\textrm{Trace}}_{\beta_m}
\biggl(D_{\beta_m}^{}(v)\prod_n^{\curvearrowright}
L_{\alpha_n,\beta_m}[\qalg_{n,m}]\biggr)
\end{equation}
may be considered.

Further in this subsection we recall the structure of $2d$ Lax
operator (\ref{Lax-L}) and its $R$-matrix (\ref{big-R}). Matrix
$\mathfrak{L}^{(n)}_{\boldsymbol{\beta}}$, as the matrix in the
auxiliary space $V_{\boldsymbol{\beta}}=(\mathbb{C}^2)^{\otimes
M}$ with operator-valued matrix elements from $\alg^{\otimes M}$,
has a block-diagonal structure. Let, in the concordance with
(\ref{44}) and (\ref{ordering}), $|0\rangle$ and $|1\rangle$ be
the basis of $\mathbb{C}^2$. Define $\varphi_0$, $\varphi_{j}$,
$\varphi_{j,j'}$, etc. as the following elements of
$V_{\boldsymbol{\beta}}$:
\begin{equation}\label{phi-basis}
\begin{array}{l}
\ds \varphi_0\;=\;|0\rangle\otimes \cdots \otimes |0\rangle\;,\\
\\
\ds \varphi_{j}\;=\;|0\rangle\otimes \cdots
\mathop{|1\rangle}_{j^{\textrm{\;th}}\textrm{ place}}\otimes
\cdots \otimes |0\rangle\;,
\end{array}
\end{equation}
etc., in general $\varphi_{j_1,...,j_m}$ has $|1\rangle$ on
$j_1^{\textrm{th}}$, ... , $j_m^{\textrm{th}}$ places,
$j_1<...<j_m$. In the basis of $\varphi$
\begin{equation}\label{L0}
\mathfrak{L}^{(n)}_{\boldsymbol{\beta}}(u)\;\varphi_0\;=\;\varphi_0\;
L_0^{(n)}(u)\;,\;\;\textrm{where}\;\;\;L_0(u)=1+u\prod_m\lambda_{n,m}
q^{\Nop_{n,m}}\;.
\end{equation}
Next
\begin{equation}
\mathfrak{L}^{(n)}_{\boldsymbol{\beta}}(u)\;\varphi_k
\;=\;\sum_{j=1}^M \varphi_j\; L^{(n)}_{j,k}(u)\;,
\end{equation}
where $L_{j,k}(u)$ are matrix elements of Lax operator for the
vector representation of $\mathcal{U}_q(\widehat{sl}_M)$, they are
given in \cite{First}. In general,
\begin{equation}
\mathfrak{L}_{\boldsymbol{\beta}}(u)\;=\;\mathop{\textrm{\large
$\oplus$}}_{m=0}^M\; L_{\omega_m}(u)\;,
\end{equation}
where block $\omega_m$ corresponds to $\frac{M!}{m!(M-m)!}$
-dimensional vector space $\pi_{\omega_m}\in
V_{\boldsymbol{\beta}}$ with the basis $\varphi_{j_1,...,j_m}$,
$j_1<\cdots<j_m$.

Matrix $\mathfrak{R}_{\boldsymbol{\beta},\boldsymbol{\beta}'}$
(\ref{big-R}) has the block structure as well,
\begin{equation}\label{big-R-dec}
\mathfrak{R}_{\boldsymbol{\beta},\boldsymbol{\beta}'}(u)\;=\;
\mathop{\textrm{\large $\oplus$}}_{m,m'=0}^M\;
\lambda_0^{m}\mu_0^{m'}R_{\omega_m^{},\omega_{m'}}(u)\;,
\end{equation}
where $\lambda_0,\mu_0$ are extra parameters of $\qalg_0$
(\ref{big-R}), and $R_{\omega_m^{},\omega_{m'}}$ is the
$\mathcal{U}_q(\widehat{sl}_M)$ $R$-matrix for the representations
$\pi_{\omega_m^{}}\otimes \pi_{\omega_{m'}}$. In particular, in
the sector $\pi_{\omega_1}\otimes \pi_{\omega_1}$ with the basis
$\varphi_j$ (\ref{phi-basis}), one can obtain the fundamental
$R$-matrix (we used the Fock space (\ref{fock}) for $\alg_0$ for
the calculation of the trace in (\ref{big-R})):
\begin{equation}
\mathfrak{R}_{\boldsymbol{\beta},\boldsymbol{\beta}'}(u)\;\varphi_j\otimes\varphi_k\;=\;
\lambda_0\mu_0 \frac{u}{(u-q^2)(u-1)}\;\sum_{j',k'}
\varphi_{j'}\otimes \varphi_{k'}\; R_{j',k'}^{j\;,k\;}(u)\;,
\end{equation}
where
\begin{equation}\label{cherednik}
\begin{array}{l}
\ds R_{j,k}^{j,k}(u)= u-1\;,\;\;\;
R_{j,j}^{j,j}=q^{-1}(u-q^2)\;,\\
\ds R_{k,j}^{j,k}(u)=q^{-1}(1-q^2)\;\;\textrm{for}\;\;j<k\;\;\;
\textrm{and}\;\;\;
R_{k,j}^{j,k}(u)=q^{-1}u(1-q^2)\;\;\textrm{for}\;\;j>k\;.
\end{array}
\end{equation}

Lax operator $\mathfrak{L}^{(n)}_{\boldsymbol{\beta}}(u)$
(\ref{Lax-L}) has the center $q^{\mathcal{J}_n}$ (\ref{charges}),
while $q^{\mathcal{K}_m}$ is the center of
$\mathfrak{L}^{(m)}_{\boldsymbol{\alpha}}(v)$ (\ref{Lax-L2}). The
quantum space of $\mathfrak{L}_{\boldsymbol{\beta}}$ is
$\mathcal{F}^{\otimes M}$, it may be decomposed as
\begin{equation}
\mathcal{F}^{\otimes M}\;=\;\mathop{\textrm{\large
$\oplus$}}_{J=0}^\infty\; \pi_{J\omega_1}\;,
\end{equation}
where $J$ is the eigenvalue of $\mathcal{J}$, and
$\pi_{J\omega_1}$ is the rank-$J$ symmetrical tensor
representation of $\mathcal{U}_q(\widehat{sl}_M)$ (dominant weight
$J\omega_1$).

We would like to conclude this subsection by the example of $M=2$
containing the six-vertex model. The block-diagonal structure of
$\mathfrak{L}$ (\ref{Lax-L}) is
\begin{equation}\label{L-for-sl2}
\mathfrak{L}(u)\;=\;\left(\begin{array}{ccc}
1+u\lambda_1\lambda_2q^{\Nop_1+\Nop_2} & 0 & 0 \\
0 & L(u) & 0 \\
0 & 0 & \mu_1\mu_2(q^{\Nop_1+\Nop_2}+q^{-2}u\lambda_1\lambda_2)
\end{array}
\right)\;,
\end{equation}
where index $n$ of (\ref{Lax-L}) is omitted, and $2\times 2$
central block is
\begin{equation}\label{2dLax}
L(u)\;=\;\left(\begin{array}{cc}
\mu_1(q^{\Nop_1}-u\lambda_1\lambda_2 q^{\Nop_2-1}) &
-q^{-1}u\lambda_1\mu_1\xop_1\yop_2\\
-q^{-1}\yop_1\lambda_2\mu_2\xop_2 &
\mu_2(q^{\Nop_2}-u\lambda_1\lambda_2q^{\Nop_1-1})
\end{array}\right)\;.
\end{equation}
Fixed integer $\mathcal{J}=\Nop_1+\Nop_2$ in the quantum space
corresponds to spin $\mathcal{J}/2$ representation of $sl_2$. For
spin $1/2$ representation ($\mathcal{J}=1$), matrix elements of
(\ref{2dLax}) may be presented by
\begin{equation}
q^{\Nop_1} = \left(\begin{array}{cc} q & 0 \\ 0 &
1\end{array}\right),\;\; q^{\Nop_2} = \left(\begin{array}{cc} 1 &
0 \\ 0 & q\end{array}\right),\;\; \xop_1\yop_2 =
\left(\begin{array}{cc} 0 & 0 \\ 1-q^2 & 0\end{array}\right),\;\;
\yop_1\xop_2 = \left(\begin{array}{cc} 0 & 1-q^2 \\ 0 &
0\end{array}\right),
\end{equation}
and in the homogeneous case
$\lambda_{1}=\lambda_{2}=\mu_{1}=\mu_{2}=1$ the matrix
(\ref{2dLax}) becomes exactly the six-vertex $R$-matrix.

Let the chain of Lax matrices (\ref{Lax-L}) is given,
$n=1,2,...,N$. Six vertex model corresponds to the choice
$\mathcal{J}_n=1$ for all $n$.  The values of two extra integrals
of motion, $\mathcal{K}_1$ and $\mathcal{K}_2$ of (\ref{charges}),
are related to total spin of the chain: $\mathcal{K}_1$ is the
number of ``spins up'', $\mathcal{K}_2$ is the number of ``spins
down''.

\section{Quantum curve}

Solution of the classical equations of motion is based on the
notion of the spectral curve $J(u,v)=0$ (\ref{determinant}). In
the previous section we succeed in construction of the ``quantum
partition function'' $\mathbf{T}(u,v)$ producing the set of the
integrals of motion, but we did not answer the question: what is
the quantum analogue of $J(u,v)=0$?

The answer is the following (we are repeating the Introduction).
Let $\uop,\vop$ be an additional auxiliary Weyl pair,
\begin{equation}\label{uv-Weyl}
\mathcal{W}\;\;:\quad \uop\;\vop\;=\;q^2\;\vop\;\uop
\end{equation}
serving two variants of ``quantum-mechanical'' notations
\begin{equation}\label{Weyl-rep}
\begin{array}{l}
\ds \langle Q| u\rangle=Q(u)\;,\quad \uop|u\rangle=|u\rangle
u\;,\quad \vop|u\rangle= |q^2u\rangle v\;,\\
\textrm{or}\\
\ds \langle v| \overline{Q}\rangle=\overline{Q}(v)\;,\quad \langle
v|\vop=v\langle v|\;,\quad \langle v|\uop = u\langle
q^2v|\;.
\end{array}
\end{equation}
For the given layer-to-layer transfer matrix $\mathbf{T}(u,v)$
(\ref{q-T-matrix},\ref{q-T-decomp}), define
\begin{equation}\label{quantum-curve}
\mathbf{J}(\uop,\vop)\;=\; \sum_{n=0}^N\sum_{m=0}^M\; (-q)^{-nm}
(-\uop)^n (-\vop)^m\;\mathbf{t}_{n,m}\;.
\end{equation}
One may easily see, in the limit $q\to 1$ (\ref{quantum-curve})
becomes (\ref{J-T-classic}). Operator $\mathbf{J}(\uop,\vop)$
belongs to $\alg^{\otimes NM}\otimes \mathcal{W}$. Let
$|\,t\rangle\in\mathcal{F}^{\otimes NM}$ be an eigenvector of
$\mathbf{T}(u,v)$,
\begin{equation}\label{t-eigenvector}
\mathbf{t}_{n,n}\;|\;t\rangle = |\;t\rangle \; t_{n,m}\;,
\end{equation}
so that
\begin{equation}
\mathbf{J}(\uop,\vop)\; \biggl( |\,t\rangle\otimes id\biggl) \;=\;
\biggl( |\,t\rangle\otimes id\biggl) \; J(\uop,\vop)\;,
\end{equation}
where $J(\uop,\vop)\in \mathcal{W}$, (\ref{J}). Define a linear
space $\Psi$ by
\begin{equation}\label{spaces}
\mathbf{J}(\uop,\vop)\;|\Psi\rangle \;=\; 0\quad \textrm{or}\quad
\langle \Psi|\;\mathbf{J}(\uop,\vop)\;=\;0\;.
\end{equation}
The linear space may be decomposed with respect to the basis
$|\,t\rangle$,
\begin{equation}
|\Psi\rangle\;=\;\mathop{\textrm{\large $\oplus$}}_t \;
|\,t\rangle \otimes |\overline{Q}_t\rangle\quad \textrm{or}\quad
\langle\Psi|\;=\;\mathop{\textrm{\large $\oplus$}}_t \; \langle t|
\otimes \langle Q_t|\;.
\end{equation}
The point is that $\langle Q_t|$ and $|\overline{Q}_t\rangle$ have
a simple and predictable structure, in particular $\langle
Q_t|u\rangle$ and $\langle v|\overline{Q}_t\rangle$ may be defined
as polynomials.

Functions $\langle Q|u\rangle$ and $\langle v|\overline{Q}\rangle$
(index $t$ is usually omitted) obey the linear difference
equations (\ref{spaces}),
\begin{equation}\label{T-Q}
\langle \Psi|\; \mathbf{J}(\uop,\vop)\; |\;t\rangle \otimes
|u\rangle \;=\;0\quad \textrm{and}\quad \langle t|\otimes\langle
v|\;\mathbf{J}(\uop,\vop)\;|\Psi\rangle \;=\;0\;,
\end{equation}
which are exactly Eqs. (\ref{BAE-slM}) and (\ref{BAE-slN}).

In this section, we will derive (\ref{quantum-curve}) and
(\ref{spaces}) as a solution of quantized linear problem
(\ref{X-kor}) for the whole auxiliary lattice. A non-trivial
problem is how to define a quantized linear problem in such a way,
that each linear variable belongs to the same subspace (subspace
(\ref{spaces}) in the final solution). This is the problem of
consistency of linear problem and invariance of quantum curve.

\subsection{Generalized matrix of coefficients of linear problem}\label{sub3.1}

Consider the linear problem (\ref{X-kor}) on an arbitrary lattice
with $u,v$-periodical boundary conditions. The set of local linear
problems
\begin{equation}\label{LPgen}
\left(\begin{array}{c} \psi_\alpha'\\
\psi_\beta'\end{array}\right)\;=\;
X_{\textrm{v}}\;\cdot\;\left(\begin{array}{c} \psi_\alpha \\
\psi_\beta\end{array}\right)\;,\;\;\;
X_\vrt\;=\;\left(\begin{array}{cc}  a_\vrt & b_\vrt \\
c_\vrt & d_\vrt
\end{array}\right)
\end{equation}
may be rewritten for the whole lattice in a matrix form,
\begin{equation}\label{all-LP}
\sum_{j}\ell_{k,j}\;\psi_j\;=\;0\;.
\end{equation}
The fragments in (\ref{all-LP}), corresponding to (\ref{LPgen}),
are
\begin{equation}\label{ell-fragment}
\ell\;=\;\left(\begin{array}{cccccc} \ddots & \vdots &
\vdots & 0 & 0 & \ddots\\
0 & 1 & 0 & -a_\vrt & -b_\vrt & 0 \\
0 & 0 & 1 & -c_\vrt & -d_\vrt & 0 \\
\ldots & 0 & 0 & 1 & 0 &\ldots\\
\ldots & 0 & 0 & 0 & 1 &\ldots\\
\ddots & \vdots & \vdots & 0 & 0 & \ddots
\end{array}
\right)\;\;\;\textrm{for}\;\;\; \psi\;=\;\left(\begin{array}{c} \vdots\\
\psi_\alpha'\\ \psi_\beta' \\ \psi_\alpha \\ \psi_\beta \\
\vdots\end{array}\right)\;.
\end{equation}
In the quantum world, matrix elements $\ell_{k,j}$ are
non-commutative operators $\boldsymbol{\ell}_{k,j}$. Therefore
(\ref{all-LP}) may have two slightly different forms,
\begin{equation}\label{two-all-LP}
\textrm{A).}\;\;\;
\sum_j\;\boldsymbol{\ell}_{k,j}\;|\psi_j\rangle=0\;\;\;\;
\textrm{or}\;\;\;\textrm{B).}\;\;\;
\sum_j\langle\psi_j|\;\boldsymbol{\ell}_{k,j}\;=\;0\;.
\end{equation}
Let the operators $\boldsymbol{\ell}_{k,j}$ obey the following
exchange relations:
\begin{equation}\label{ell-relations}
\begin{array}{l}
\ds
\boldsymbol{\ell}_{k,j}\boldsymbol{\ell}_{k',j'}-\boldsymbol{\ell}_{k',j}\boldsymbol{\ell}_{k,j'}
\;=\;
\boldsymbol{\ell}_{k',j'}\boldsymbol{\ell}_{k,j}-\boldsymbol{\ell}_{k,j'}\boldsymbol{\ell}_{k',j}\;,\;\;\;
k\neq
k'\;,\;\;\; j\neq j'\;, \\
\ds \textrm{and} \\
\ds
\boldsymbol{\ell}_{k,j}\boldsymbol{\ell}_{k',j}=\boldsymbol{\ell}_{k',j}\boldsymbol{\ell}_{k,j}\;.
\end{array}
\end{equation}
The aim of this subsection is to establish fundamental properties
of such $\boldsymbol{\ell}$ and the form of solution of
(\ref{two-all-LP}) A) and B).

Define
\begin{equation}\label{ell-det}
\det\boldsymbol{\ell}\;\stackrel{\textrm{def}}{=}\; \sum_\sigma
(-)^\sigma \prod_j^\curvearrowright
\boldsymbol{\ell}_{\sigma_j,j}\;\equiv\; \sum_{\sigma} (-)^\sigma
\boldsymbol{\ell}_{\sigma_1,1}\boldsymbol{\ell}_{\sigma_2,2}\boldsymbol{\ell}_{\sigma_3,3}\;\dots
\end{equation}
where $\sigma$ are the permutations of the indices
$1,2,3,\dots$\;. According to the first relation of
(\ref{ell-relations}), the definition (\ref{ell-det}) is invariant
with respect to the ordering of $j$, for instance
\begin{equation}
\sum_\sigma (-)^\sigma
\boldsymbol{\ell}_{\sigma_1,1}\boldsymbol{\ell}_{\sigma_2,2}\boldsymbol{\ell}_{\sigma_3,3}\;\dots
= \sum_\sigma (-)^\sigma
\boldsymbol{\ell}_{\sigma_2,2}\boldsymbol{\ell}_{\sigma_1,2}\boldsymbol{\ell}_{\sigma_3,3}\;\dots
=\dots,
\end{equation}
and in general
\begin{equation}\label{det-anypermut}
\det\boldsymbol{\ell}\;=\;\sum_\sigma (-)^\sigma
\prod_j^\curvearrowright
\boldsymbol{\ell}_{\sigma_{\tau_j},\tau_j}
\end{equation}
where $\tau$ is any permutation of $1,2,3,\dots$. Define next the
algebraic supplements $\mathbf{A}_{j,k}$ of $\boldsymbol{\ell}$
(adjoint matrix) as (\ref{ell-det})-determinants of the minors of
$\boldsymbol{\ell}$,
\begin{equation}
\sum_k \mathbf{A}_{j,k}\;\boldsymbol{\ell}_{k,j}\;=\;\sum_k
\boldsymbol{\ell}_{k,j} \;
\mathbf{A}_{j,k}\;=\;\det\boldsymbol{\ell}\;.
\end{equation}
The first equality here follows from (\ref{det-anypermut}), the
last one in the definition of $\mathbf{A}_{j,k}$. The second line
of (\ref{ell-relations}) provides
\begin{equation}
\sum_k
\mathbf{A}_{j,k}\;\boldsymbol{\ell}_{k,j'}\;=\;\sum_k\boldsymbol{\ell}_{k,j'}\;
\mathbf{A}_{j,k}\;=\; 0\;\;\;\textrm{if}\;\;\;j\neq j'\;.
\end{equation}
Therefore $(\det\boldsymbol{\ell})^{-1}\mathbf{A}_{j,k}$ and
$\mathbf{A}_{j,k}(\det\boldsymbol{\ell})^{-1}$ are two variants of
inverse matrices,
\begin{equation}
\sum_k (\det\boldsymbol{\ell})^{-1} \mathbf{A}_{j,k}
\boldsymbol{\ell}_{k,j'}\;=\;\sum_k \boldsymbol{\ell}_{k,j'}
\mathbf{A}_{j,k} (\det\boldsymbol{\ell})^{-1}\;=\;\delta_{j,j'}\;,
\end{equation}
or, since the inverse matrix must be both left- and right-inverse,
\begin{equation}\label{inverse-ell}
\sum_j \boldsymbol{\ell}_{k,j}(\det\boldsymbol{\ell})^{-1}
\mathbf{A}_{j,k_0}\;=\; \sum_j \mathbf{A}_{j,k_0}
(\det\boldsymbol{\ell})^{-1}\boldsymbol{\ell}_{k,j}\;=\;
\delta_{k,k_0}\;.
\end{equation}
The main property of the elements of inverse matrices is the
commutativity of their matrix elements with the same $k_0$, e.g.
for the variant A),
\begin{equation}\label{inverse-commut}
\left[ (\det\boldsymbol{\ell})^{-1}
\mathbf{A}_{j,k_0}\;,\;(\det\boldsymbol{\ell})^{-1}
\mathbf{A}_{j',k_0}\right]\;=\;0\;.
\end{equation}
It is a particular case of the following statement:
\begin{equation}\label{ell-m-epsilon}
\sum_j \;\boldsymbol{\ell}_{k,j}\; \mathbf{m}_j \;=\;
\varepsilon_k\;,\;\;\;\varepsilon_k\in\mathbb{C}\;\;\;
\Rightarrow\;\;\; [\mathbf{m}_j,\mathbf{m}_{j'}]\;=\;0\;.
\end{equation}
To prove (\ref{ell-m-epsilon}), consider
\begin{equation}
c_{k,k'}=\sum_{j<j'}
(\boldsymbol{\ell}_{k,j}\boldsymbol{\ell}_{k',j'}-\boldsymbol{\ell}_{k',j
}\boldsymbol{\ell}_{k,j'}) (\mathbf{m}_j \mathbf{m}_{j'} -
\mathbf{m}_{j'} \mathbf{m}_j)\;,\;\;\;k<k'\;.
\end{equation}
Due to the first relation of (\ref{ell-relations}) and definition
(\ref{ell-m-epsilon}), $c_{k,k'} =
\varepsilon_k\varepsilon_{k'}-\varepsilon_{k'}\varepsilon_k = 0$.
From the other side, let $\mathbf{A}^{(2)}_{j,j'|k,k'}$ is the
matrix of the second algebraical supplements of
$\boldsymbol{\ell}$:
\begin{equation}
\sum_{k<k'} \mathbf{A}^{(2)}_{i,i'|k,k'}
(\boldsymbol{\ell}_{k,j}\boldsymbol{\ell}_{k',j'}-\boldsymbol{\ell}_{k',j
}\boldsymbol{\ell}_{k,j'})\;=\;\delta_{i,j}\delta_{i',j'}\;\det\boldsymbol{\ell}\;,\;\;\;
i<i'\;,\;\;j<j'\;.
\end{equation}
Then $\ds 0 \;=\; \sum_{k<k'} \mathbf{A}^{(2)}_{i,i'|k,k'}
c_{k,k'}= \det\boldsymbol{\ell}\; \,\cdot\,(\mathbf{m}_j
\mathbf{m}_{j'} - \mathbf{m}_{j'} \mathbf{m}_j)$, what proves the
commutativity of $\mathbf{m}_j$. The commutativity
(\ref{inverse-commut}) corresponds to
$\varepsilon_k=\delta_{k,k_0}$.

For $\varepsilon_k=\delta_{k,k_0}$ with fixed $k_0$, let
\begin{equation}\label{q-BA}
\mathbf{m}^{}_{j,j_0}\;=\;(\mathbf{m}_{j_0})^{-1}\mathbf{m}_j\;=\;\mathbf{A}_{j_0,k_0}^{-1}
\mathbf{A}_{j,k_0}^{}\;,\;\;\;
\mathbf{m}'_{j,j_0}=\mathbf{A}_{j,k_0}^{}
\mathbf{A}_{j_0,k_0}^{-1}\;.
\end{equation}
Then the set $\mathbf{m}_{j,j_0}$ is commutative ($k_0$ is fixed),
as well the set $\mathbf{m}'_{j,j_0}$, and besides
\begin{equation}\label{mmp}
\det\boldsymbol{\ell}\,\cdot\,\mathbf{m}^{}_{j,j_0}\;=\;
\mathbf{m}'_{j,j_0}\,\cdot\,\det\boldsymbol{\ell}\;.
\end{equation}
Inverse relations (\ref{inverse-ell}) give
\begin{equation}\label{sol-LP}
\sum_{j} \boldsymbol{\ell}_{k,j}
\mathbf{m}^{}_{j,j_0}\;=\;\delta_{k,k_0}
\mathbf{A}_{j_0,k_0}^{-1}\;\det\boldsymbol{\ell}\;,\;\;\;\; \sum_j
\mathbf{m}_{j,j_0}'\;\boldsymbol{\ell}_{k,j}\;=\; \delta_{k,k_0}
\det\boldsymbol{\ell} \; \mathbf{A}_{j_0,k_0}^{-1}\;.
\end{equation}
Equations (\ref{sol-LP}) give the formal solution of the linear
problem (\ref{two-all-LP}) A) and B): let $|\psi_{j_0}\rangle$ or
$\langle\psi_{j_0}|$ for some $j_0$ are defined by
$\det\boldsymbol{\ell}\;\cdot\;|\psi_{j_0}\rangle=0$ or
$\langle\psi_{j_0}|\;\cdot\;\det\boldsymbol{\ell}\;=\;0$. Then
\begin{equation}\label{mm}
\textrm{A).}\;\;\;|\psi_j\rangle\;=\;\mathbf{m}_{j,j_0}|\psi_{j_0}
\rangle\;\;\;\textrm{or}\;\;\;
\textrm{B).}\;\;\;\langle\psi_j|\;=\;\langle\psi_{j_0}|\mathbf{m}_{j,j_0}'\;.
\end{equation}
Common feature of all $|\psi_j\rangle$ or $\langle\psi_j|$ is
\begin{equation}\label{all-j}
\textrm{A).}\;\;\;\det\boldsymbol{\ell}\;\cdot\;|\psi_j\rangle\;=\;0\quad\textrm{or}\quad
\textrm{B).}\;\;\;\langle\psi_j|\;\cdot\;\det\boldsymbol{\ell}\;=\;0\;\;\;\forall\;\;j\;.
\end{equation}

\subsection{Extended algebra of observables}\label{sub3.2}

To establish the relation between the algebra of matrix elements
(\ref{ell-relations}) and our case of $X$ (\ref{quantum-X}), we
have to introduce the extended algebra of observables
$\mathfrak{A}_\vrt=\alg_\vrt\otimes \mathcal{W}_\vrt$. Define
\begin{equation}\label{113}
X[\mathfrak{A}_\vrt]\;=\;\left(\begin{array}{cc}
\ds \boldsymbol{\lambda}_\vrt q^{\Nop_\vrt} & \ds \boldsymbol{\nu}_\vrt\yop_\vrt\\
\ds \boldsymbol{\nu}_\vrt\xop_\vrt & \ds \boldsymbol{\mu}\vrt
q^{\Nop_\vrt}\end{array} \right)\;,
\end{equation}
where $q^{\Nop_\vrt},\xop_\vrt,\yop_\vrt$ are generators of
$q$-oscillator $\alg_\vrt$ (\ref{q-osc}) and
$\boldsymbol{\lambda}_\vrt,\boldsymbol{\mu}_\vrt,\boldsymbol{\nu}_\vrt$
are the generators of Weyl algebra $\mathcal{W}_\vrt$:
\begin{equation}\label{lm-Weyl}
\boldsymbol{\lambda}_\vrt\boldsymbol{\mu}_\vrt\;=\;
q^2\boldsymbol{\mu}_\vrt\boldsymbol{\lambda}_\vrt\;,\;\;\;
\boldsymbol{\nu}_\vrt^2\;=\;-\boldsymbol{\mu}_\vrt^{}\boldsymbol{\lambda}_\vrt^{}\;.
\end{equation}
The $q$-oscillators and the Weyl algebra elements for different
vertices always commute.

Matrix $\boldsymbol{\ell}$, defined according to
(\ref{ell-fragment}) for vertex matrix $X[\mathfrak{A}_\vrt]$
(\ref{113}), belongs exactly to the class (\ref{ell-relations})
since
\begin{equation}
\begin{array}{l}
\ds \boldsymbol{\nu}\xop\;\boldsymbol{\nu}\yop -
\boldsymbol{\lambda} q^\Nop\;\boldsymbol{\mu} q^\Nop \;=\;
\boldsymbol{\nu}\yop\;\boldsymbol{\nu}\xop - \boldsymbol{\mu}
q^\Nop\;\boldsymbol{\lambda} q^\Nop\;,\\
\ds\textrm{and}\\
\ds [\boldsymbol{\nu}\yop\;,\;\boldsymbol{\mu}q^\Nop] \;=\;
[\boldsymbol{\nu}\xop\;,\;\boldsymbol{\lambda}q^\Nop] \;=\; 0\;
\end{array}
\end{equation}
and elements of $\mathfrak{A}_\vrt$ for different vertices
commute. Determinant of $\boldsymbol{\ell}$ has a combinatorial
representation of Fig. \ref{fig-fragments}, eqs.
(\ref{t_C},\ref{t_nm},\ref{J-T-classic}) with
\begin{equation}\label{113a}
L_{\alpha,\beta}[\mathfrak{A}_\vrt]\;=\;\left(\begin{array}{cccc} 1 & 0 & 0 & 0 \\
0 & \boldsymbol{\lambda}_\vrt q^{\Nop_{\vrt}} & \boldsymbol{\nu}_\vrt\yop_\vrt & 0 \\
0 & \boldsymbol{\nu}_\vrt\xop_\vrt & \boldsymbol{\mu}_\vrt q^{\Nop_\vrt} & 0 \\
0 & 0 & 0 & \boldsymbol{\nu}^2_\vrt \end{array}\right)\;.
\end{equation}
Let
\begin{equation}\label{detell}
\det\boldsymbol{\ell}\;\stackrel{\textrm{def}}{=}\;\mathbf{J}[\mathfrak{A}]\;=\;
\sum (-)^{nm+n+m} u^n v^m \mathbf{J}_{n,m}[\mathfrak{A}]\;,
\end{equation}
where $u$ and $v$ are $\mathbb{C}$-valued spectral parameters
introduced according to (\ref{uv-bc}). Let locally
\begin{equation}
\boldsymbol{\lambda}_\vrt\;=\;\lambda_\vrt \EXP^{Q_\vrt}\;, \;\;\;
\boldsymbol{\mu}_\vrt\;=\;\mu_\vrt \EXP^{P_\vrt}\;\;\textrm{and
therefore}\;\;
\boldsymbol{\nu}^2_\vrt\;=\;-q^{-1}\lambda_\vrt\mu_\vrt
\EXP^{P_\vrt+Q_\vrt}\;,\;\;\;[Q_\vrt,P_\vrt]=\log q^2\;.
\end{equation}
In the combinatorial representation of the determinant (Fig.
\ref{fig-fragments}, eqs.
(\ref{t_C},\ref{t_nm},\ref{J-T-classic})), consider a path
$C_{n,m}$ of the homotopy class $n\mathcal{A}+m\mathcal{B}$. A
monomial summand, corresponding to this path, may be factorized as
\begin{equation}
\mathbf{J}_{C_{n,m}}\;=\; \mathbf{t}_{C_{n,m}}\;
\EXP^{\boldsymbol{\phi}(C_{n,m})}\;.
\end{equation}
Monomial $\mathbf{t}_{C_{n,m}}$ gathers all the $q$-oscillators
and $\mathbb{C}$-valued parameters $\lambda_\vrt,\mu_\vrt$ and
$-q^{-1}\lambda_\vrt\mu_\vrt$, and therefore it is exactly the
$C_{n,m}$-monomial of $\mathbf{T}(u,v)$ (up to unessential
renormalization of $\xop_\vrt$ and $\yop_\vrt$). Operator
$\boldsymbol{\phi}$ is a sum of local $Q_\vrt$ and $P_\vrt$. One
may easily see,
\begin{equation}\label{f-alg}
[\boldsymbol{\phi}(C_{n,m}),\boldsymbol{\phi}(C'_{n',m'})]\ \;=\;
(nm'-mn')\log q^2\;\;\;\;\forall\;\;C,C'\;.
\end{equation}
Let $C_{1,0}$ and $C_{0,1}$ be two particular fixed paths. Due to
(\ref{f-alg}), all
\begin{equation}\label{Weyl-centers}
\widetilde{\boldsymbol{\phi}}(C_{n,m})\;=\;
\boldsymbol{\phi}(C_{n,m})- n\boldsymbol{\phi}(C_{1,0})-
m\boldsymbol{\phi}(C_{0,1})
\end{equation}
commute. Therefore one may diagonalize them simultaneously and
without lost of generality (since $\lambda_\vrt,\mu_\vrt$ are
free) put
\begin{equation}\label{restriction}
\widetilde{\phi}(C_{n,m})\;\equiv\;0\;.
\end{equation}
On this condition,
\begin{equation}
\boldsymbol{\phi}(C_{n,m})\;\to\;n\boldsymbol{\phi}(C_{1,0})+
m\boldsymbol{\phi}(C_{0,1})\;\equiv\;
nQ_0+mP_0\;,
\end{equation}
and exponent of $\boldsymbol{\phi}(C_{n,m})$, together with the
combinatorial factor $(-)^{nm+n+m}u^nv^m$, becomes
\begin{equation}\label{Weyl-factor}
(-)^{nm+n+m} \; \EXP^{nQ_0+mP_0}\; u^n v^m\;=\; (-q)^{-nm}
(-\uop)^n (-\vop)^m\;,\;\;\;
\uop=\EXP^{Q_0}u\;,\;\;\vop=\EXP^{P_0} v\;.
\end{equation}
Therefore, on the subspace (\ref{restriction}),
$\mathbf{J}[\mathfrak{A}]$ becomes exactly $\mathbf{J}(\uop,\vop)$
(\ref{quantum-curve}).

Definition of $\uop,\vop$ (\ref{Weyl-factor}) on the subspace
(\ref{restriction}) may be written in the terms of
$\boldsymbol{\lambda}_\vrt$, $\boldsymbol{\mu}_\vrt$ as
\begin{equation}\label{uop-vop}
u\prod_{m}\boldsymbol{\lambda}_{nm}\;
\stackrel{\widetilde{\phi}=0}{\mapsto} \;
\uop\prod_m\lambda_{nm}\;,\;\;\; v\prod_n\boldsymbol{\mu}_{nm}
\stackrel{\widetilde{\phi}=0}{\mapsto} \; \vop\prod_n\mu_{nm}\;.
\end{equation}

\subsection{Structure of a solution of linear problem}

Turn now to the structure of linear spaces $|\Psi\rangle$ and
$\langle \Psi|$ defined by
\begin{equation}\label{inv-spaces}
\textrm{A).}\;\;\; \mathbf{J}[\mathfrak{A}]\;|\Psi\rangle\;=\;0
\;\;\;\textrm{and}\;\;\; \textrm{B).} \;\;\;
\langle\Psi|\;\mathbf{J}[\mathfrak{A}]\;=\;0\;.
\end{equation}
According to (\ref{inverse-commut},\ref{q-BA},\ref{mmp}), each of
$|\psi_j\rangle$ and $\langle\psi_j|$ belong to these subspaces,
and there exist sets of commutative operators $\mathbf{m}_j$ and
$\mathbf{m}_j'$ (\ref{mm}) such that
$|\psi_j\rangle=\mathbf{m}_j\;|\Psi\rangle$ and
$\langle\psi_j|=\langle\Psi|\;\mathbf{m}'_j$. In particular, one
may consider the linear spaces $|\Psi\rangle$ and $\langle\Psi|$
in the basis of diagonal $\mathbf{m}_j,\mathbf{m}_j'$:
\begin{equation}\label{separated}
\textrm{A).}\;\;\;|\psi_j\rangle \;=\; |\Psi\rangle m_j\;,\quad
\textrm{or}\quad \textrm{B).}\;\;\;
\langle\psi_j|=m_j\langle\Psi|\;,
\end{equation}
where $m_j$ are eigenvalues of $\mathbf{m}_j$ etc. The every local
pair of linear equations (\ref{LPgen}) take the form
\begin{equation}\label{local-inv-space}
\textrm{A).}\;\;\left\{\begin{array}{l} \ds
(1-\boldsymbol{\lambda}_\vrt q^{\Nop_\vrt} A_\vrt -
\boldsymbol{\nu}_\vrt \yop_\vrt B_\vrt)|\Psi\rangle=0\\ \\ \ds
(1-\boldsymbol{\nu}_\vrt\xop_\vrt C_\vrt - \boldsymbol{\mu}_\vrt
q^{\Nop_\vrt} D_\vrt)|\Psi\rangle=0
\end{array}\right.\;\;\;\textrm{or}\;\;\; \textrm{B).}\;\;
\left\{\begin{array}{l} \ds
\langle\Psi|(1-A_\vrt\boldsymbol{\lambda}_\vrt q^{\Nop_\vrt}  -
B_\vrt \boldsymbol{\nu}_\vrt \yop_\vrt)=0\\ \\ \ds \langle\Psi|(1-
C_\vrt \boldsymbol{\nu}_\vrt\xop_\vrt  - D_\vrt
\boldsymbol{\mu}_\vrt q^{\Nop_\vrt} )=0
\end{array}\right.
\end{equation}
where in the notations of (\ref{LPgen}) $\ds
A_\vrt=\frac{m_\alpha}{m_{\alpha'}}$, $\ds
B_\vrt=\frac{m_\beta}{m_{\alpha'}}$, $\ds
C_\vrt=\frac{m_\alpha}{m_{\beta'}}$ and $\ds
D_\vrt=\frac{m_\beta}{m_{\beta'}}$.

The linear spaces $|\Psi\rangle$ and $\langle\Psi|$ belong to a
module of $\alg^{\otimes NM}\otimes\mathcal{W}^{\otimes NM}$. Its
``physical'' part is $\mathcal{F}^{\otimes NM}$, the artificial
Weyl part may be defined in many inequivalent ways. Let
$|n\rangle$ be the state of $\mathcal{F}^{\otimes NM}$ with
$n_\vrt$ bosons in the vertex $\vrt$, $n\;=\;\{n_\vrt\}$. We will
focus on the formal states of $\mathcal{W}^{\otimes NM}$ for
(\ref{inv-spaces}) in the basis (\ref{separated}):
\begin{equation}\label{semi-element}
\textrm{A).}\;\;\;|\Psi_{n}\rangle \;=\; \biggl(\langle n|\otimes
id\biggr) |\Psi\rangle \quad \textrm{or}\quad
\langle\Psi_n|\;=\;\langle\Psi| \biggl( |n\rangle \otimes
id\biggr) \;.
\end{equation}
Equations (\ref{local-inv-space}) give for (\ref{semi-element})
\begin{equation}\label{local-inv-solution}
\begin{array}{l}
\ds \textrm{A).}\;\;\;\left\{\begin{array}{l} \ds
(1-\boldsymbol{\lambda}_\vrt A_\vrt)|\Psi_0\rangle=0\quad \forall\;\vrt\quad \textrm{and}
\\
\ds |\Psi_n\rangle = \prod_\vrt
\frac{\boldsymbol{\nu}_\vrt^{-n_\vrt}(\boldsymbol{\mu}_\vrt
D_\vrt;q^2)_n}{C_\vrt^{n_\vrt}\sqrt{(q^2;q^2)_{n_\vrt}}}\;
|\Psi_0\rangle
\end{array}
\right.\\
\textrm{or}\\
\textrm{B).}\;\;\; \left\{\begin{array}{l} \ds
\langle\Psi_0|\;(1-\boldsymbol{\mu}_\vrt D_\vrt)\;=\;0\quad \forall \; \vrt \quad \textrm{and} \\
\ds \langle\Psi_n| = \langle\Psi_0|\;\prod_\vrt
\frac{(\boldsymbol{\lambda}_\vrt A_\vrt ;q^2)_{n_\vrt}
\boldsymbol{\nu}_\vrt^{-n_\vrt}}{B_\vrt^{n_\vrt}\sqrt{(q^2;q^2)_{n_\vrt}}}
\end{array}
\right.
\end{array}
\end{equation}
Here we used the representation
\begin{equation}
\xop\;|n\rangle\;=\;|n-1\rangle \;\sqrt{1-q^{2n}}\;,\;\;\;
\yop\;|n\rangle\;=\;|n+1\rangle\;\sqrt{1-q^{2+2n}}\;.
\end{equation}

Turn at the first to the conditions for $\Psi_0$ in
(\ref{local-inv-solution}). Since the total Fock vacuum is the
eigenstate of $q$-oscillator counterpart of
$\mathbf{J}[\mathfrak{A}]$, one may consider directly
\begin{equation}
\begin{array}{l}
\ds \textrm{A).}\;\;\; 0\;=\;\biggl(\langle 0|\otimes id\biggr) \;
\mathbf{J}[\mathfrak{A}]\;|\Psi\rangle \;=\;
\mathbf{J}_0[\mathcal{W}^{\otimes NM}]\; |\Psi_0\rangle\\
\textrm{or}\\
\ds \textrm{B).}\;\;\; 0\;=\;\langle\Psi| \;
\mathbf{J}[\mathfrak{A}]\; \biggl(|0\rangle \otimes id\biggr)
\;=\; \langle\Psi_0|\; \mathbf{J}_0[\mathcal{W}^{\otimes NM}]\;,
\end{array}
\end{equation}
where
\begin{equation}
\mathbf{J}_0[\mathcal{W}^{\otimes NM}]\;=\;\prod_m(1-v\prod_n
\boldsymbol{\mu}_{nm})\prod_n(1-u\prod_m\boldsymbol{\lambda}_{nm})\;.
\end{equation}
Evidently, $\mathbf{J}_0$ commutes with all
$\widetilde{\boldsymbol{\phi}}$ (\ref{Weyl-centers}) and therefore
allows the projection $\mathcal{W}^{\otimes NM}\to \mathcal{W}$
(\ref{restriction}). Simply applying (\ref{uop-vop}),
\begin{equation}
\mathbf{J}_0[\mathcal{W}^{\otimes NM}]
\stackrel{\widetilde{\phi}=0}{\mapsto} J_0(\uop,\vop) \;=\;
\prod_m(1-\vop\prod_n
\mu_{nm})\prod_n(1-\uop\prod_m\lambda_{nm})\;.
\end{equation}
Now the state of the Weyl algebra $\uop,\vop$ may be chosen in the
most convenient way. For the case A the proper basis is $\langle
v|$ of (\ref{Weyl-rep}), and for the case B the proper basis is
$|u\rangle$. If $u=u_n$ for the case A or $v=v_m$ for the case B,
\begin{equation}\label{un-vm}
u_n\;=\;\prod_m \lambda_{nm}^{-1}\;,\quad v_m =
\prod_n\mu_{nm}^{-1}\;,
\end{equation}
then
\begin{equation}\label{vacuum-Psi}
\textrm{A).}\;\;\;\langle v|\Psi_0\rangle \;=\; 1\quad
\textrm{or}\quad \textrm{B).} \;\;\; \langle \Psi_0|u\rangle
\;=\;1\;.
\end{equation}

Let further $|t\rangle$ and $\langle t|$ be the eigenstates of
$\mathbf{t}_{nm}$, (\ref{t-eigenvector}). Analogously to
(\ref{semi-element}) let
\begin{equation}\label{semi-element2}
\textrm{A).}\;\;\;|\Psi_{t}\rangle \;=\; \biggl(\langle t|\otimes
id\biggr) |\Psi\rangle \quad \textrm{or}\quad
\textrm{B).}\;\;\;\langle\Psi_t|\;=\;\langle\Psi| \biggl(
|t\rangle \otimes id\biggr) \;.
\end{equation}
It follows from the second lines of (\ref{local-inv-solution})
\begin{equation}
\textrm{A).}\;\;\; |\Psi_t\rangle \;=\;
\mathcal{P}_t(\boldsymbol{\nu},\boldsymbol{\mu}) \; |\Psi_0\rangle
\quad \textrm{or}\quad \textrm{B).}\;\;\;
\langle\Psi_t|\;=\;\langle\Psi_0|\;\mathcal{P}'_t(\boldsymbol{\nu},\boldsymbol{\lambda})
\;,
\end{equation}
where e.g. $\mathcal{P}_t$ is a polynomial of
$\boldsymbol{\mu}_\vrt$ of a total power not more that the number
of bosons in $|t\rangle$, its structure with respect to
$\boldsymbol{\nu}_\vrt$ is rather simple. In addition, since we
consider the eigenstates in the Fock counterpart of
$\mathfrak{A}^{\otimes NM}$, polynomials $\mathcal{P}_t$ and
$\mathcal{P}'_t$ must commute with all
$\widetilde{\boldsymbol{\phi}}$ (\ref{Weyl-centers}) and therefore
must allow the projection $\mathcal{W}^{\otimes NM}\to
\mathcal{W}$ (\ref{restriction}):
\begin{equation}
\mathcal{P}_t \to \wop^{-J} P_t(\vop)\quad \textrm{and}\quad
\mathcal{P}'_t\to P'_t(\uop) \wop^{-K}\;,
\end{equation}
where $\wop^2=-\vop\uop$, $\wop$-factor comes from
$\boldsymbol{\nu}_\vrt$ factors; $J$ and $K$ are integers; $P_t$
and $P'_t$ are polynomials of a power not higher than the total
number of bosons in the state $|\,t\rangle$.

Taking now into account (\ref{vacuum-Psi}), we come to the final
statement: in the restricted algebra $\alg^{\otimes NM}\otimes
\mathcal{W}$ the equations
\begin{equation}\label{inv-spaces-uv}
\textrm{A).} \;\;\; \biggl(\langle t|\otimes \langle v|\biggr)
\;\mathbf{J}(\uop,\vop)\;|\Psi\rangle\;=\;0\quad
\textrm{and}\;\;\;
\textrm{B).}\;\;\;\langle\Psi|\;\mathbf{J}(\uop,\vop)\;\biggl(|t\rangle\otimes
|u\rangle\biggr)\;=\;0\;.
\end{equation}
have the solutions
\begin{equation}
\begin{array}{l}
\ds \textrm{A).}\;\;\; \textrm{if}\;\; u=u_n
\;\;\textrm{then}\;\;\;\langle v|\Psi\rangle \;=\; v^{-J_n/2}
\overline{Q}_n(v)\\
\textrm{and}\\
\ds \textrm{B).}\;\;\; \textrm{if}\;\; v=v_m \;\; \textrm{then}
\;\;\; \langle\Psi|u\rangle \;=\; u^{-K_m/2} Q_m(u)
\end{array}
\end{equation}
where $u_n,v_m$ are given by (\ref{un-vm}), and the integers
$J_n$, $K_m$ and degrees of polynomials $Q_n$, $\overline{Q}_m$
are not higher that the total number of bosons.

In the next section, considering the examples, we will see that
$J_n$ and $K_m$ are the eigenvalues of $\mathcal{J}_n$ and
$\mathcal{K}_m$ (\ref{charges}) and, moreover, the degrees of
$\overline{Q}_m$ and $Q_n$ are exactly $K_m$ and $J_n$.

In the next section we will add $q^{-\mathcal{J}_n}$ and
$q^{-\mathcal{K}_m}$ to the definition of $u_n$ and $v_m$. This
allows one to cancel the half-integer pre-factors $v^{-J_n/2}$ of
$\overline{Q}_n(v)$ and $u^{-K_m/2}$ of $Q_m(u)$. The
corresponding values of $u_n$ and $v_m$ may be obtained via
conditions $\overline{Q}(0)=1$ and $Q_n(0)=1$.

\section{Examples}

Let us illustrate (\ref{T-Q}) for the six-vertex model first, and
then for an arbitrary square lattice.

\subsection{Six-vertex chain}

Consider the lattice with $M=2$ and arbitrary $N$.  This is the
case of six-vertex model, where (\ref{T-Q}) becomes the Baxter
equation. It follows from (\ref{L-for-sl2}),
\begin{equation}\label{t012}
\begin{array}{l}
\ds \sum_n u^n \mathbf{t}_{n,0}\;=\;\prod_n
\left(1+u\lambda_{n,1}\lambda_{n,2}
q^{\Nop_{n,1}+\Nop_{n,2}}\right)\;,\\
\\
\ds \sum_n u^n \mathbf{t}_{n,1}\;=\;\mathbf{t}(u)\;,\\
\\
\ds \sum_n u^n \mathbf{t}_{n,2}\;=\;\prod_n \mu_{n,1}\mu_{n,2}
\left(q^{\Nop_{n,1}+\Nop_{n,2}}+q^{-2} u
\lambda_{n,1}\lambda_{n,2}\right)\;,
\end{array}
\end{equation}
where $\mathbf{t}(u)$ is the transfer matrix for the Lax operator
(\ref{2dLax}). Its opposite elements are
\begin{equation}\label{t0-tN}
\begin{array}{l}
\ds \mathbf{t}_{0,1}\;=\;\prod_n \left(\mu_{n,1}
q^{\Nop_{n,1}}\right)
\;+\; \prod_n \left(\mu_{n,2} q^{\Nop_{n,2}}\right)\;,\\
\\
\ds \mathbf{t}_{N,1}\;=\; \prod_n
\left(-q^{-1}\lambda_{n,1}\lambda_{n,2}\mu_{n,2}
q^{\Nop_{n,1}}\right) \;+\; \prod_n
\left(-q^{-1}\lambda_{n,2}\lambda_{n,1}\mu_{n,1}
q^{\Nop_{n,2}}\right)\;.
\end{array}
\end{equation}
Applying the rule (\ref{quantum-curve})
$u^nv^m\;\mapsto\;(-q)^{-nm}(-\uop)^n (-\vop)^m$, we come to
\begin{equation}
\begin{array}{l}
\ds \mathbf{J}(q\uop,\vop)\vop^{-1}\;=\; -\mathbf{t}(\uop)\\
\\
\ds  +\vop^{-1}\prod_n\left(1-q^{-1}\uop\lambda_{n,1}\lambda_{n,2}
q^{\Nop_{n,1}+\Nop_{n,2}}\right)
\ds + \vop\prod_n
\mu_{n,1}\mu_{n,2}\left(q^{\Nop_{n,1}+\Nop_{n,2}}-q^{-1}\uop\lambda_{n,1}\lambda_{n,2}\right)\;.
\end{array}
\end{equation}
Equation $\langle Q|J(q\uop,\vop)\vop^{-1}|u\rangle=0$ is exactly
Baxter's equation for $\mathcal{U}_q(\widehat{sl}_2)$,
\begin{equation}\label{tq-example}
Q(u) t(u) \;=\; Q(q^{-2}u) \phi(u) \;+\; Q(q^2u) \phi'(u)
\end{equation}
where
\begin{equation}
\begin{array}{l}
\ds \phi(u)\;=\;v^{-1}\prod_n
\left(1-q^{-1}u\lambda_{n,1}\lambda_{n,2}q^{\Nop_{n,1}+\Nop_{n,2}}\right)\;,\\
\\ \textrm{and}\\
\ds \phi'(u)\;=\;v\prod_{n} \mu_{n,1}\mu_{n,2}
\left(q^{\Nop_{n,1}+\Nop_{n,2}} -
q^{-1}u\lambda_{n,1}\lambda_{n,2}\right)\;.
\end{array}
\end{equation}
Condition $Q(0)=1$ gives (see (\ref{t0-tN}))
\begin{equation}
v^{-1} + v\prod_{n}\mu_{n,1}\mu_{n,2}q^{\Nop_{n,1}+\Nop_{n,2}}
\;=\; \prod_n\mu_{n,1}q^{\Nop_{n,1}}\;+\;
\prod_n\mu_{n,2}q^{\Nop_{n,2}}\;,
\end{equation}
which has two solutions corresponding to two Baxter's functions
$Q$: the first one is
\begin{equation}
v=v_1\stackrel{\textrm{def}}{=}\left(\prod_n
\mu_{n,1}q^{\Nop_{n,1}}\right)^{-1}\;,
\end{equation}
so that $Q=Q_1(u)$ is a polynomial of the degree $\ds \sum_n
\Nop_{n,1}\;=\;\mathcal{K}_1$ (see (\ref{charges})), this value of
the degree follows from the second line of eq. (\ref{t0-tN}); the
second solution is
\begin{equation}
v=v_2\stackrel{\textrm{def}}{=}\left(\prod_n
\mu_{n,2}q^{\Nop_{n,2}}\right)^{-1}\;,
\end{equation}
and corresponding $Q=Q_2(u)$ is the polynomial of the power $\ds
\sum_n \Nop_{n,2}\;=\;\mathcal{K}_2$. In some sense, two functions
$Q$ correspond to two sheets of classical $q=1$ spectral
hyperelliptic curve $v^{-1}\phi(u)+v\phi'(u)=t(u)$. Note, we
consider now the Bethe Ansatz equations for
$\mathcal{U}_q(\widehat{sl}_2)$ chain with arbitrary
$\mathcal{J}_n$, this is the inhomogeneity of highest spin.
Six-vertex case corresponds to $\mathcal{J}_n=1$ for all $n$.

In the \emph{inhomogeneous} case the spectrum of $t(u)$ follows
from (\ref{tq-example}) and just the condition $Q\neq 0$ (without
fixing $v$ and considering the polynomial structure of $Q$). Let
us fix $\mathcal{J}_n=1$, the six-vertex case and a priori the
Fock space representation. Simply substituting
$u=(\lambda_{n,1}\lambda_{n,2})^{-1}$ and
$u=q^2(\lambda_{n,1}\lambda_{n,2})^{-1}$ into (\ref{tq-example})
and excluding $Q$, one comes to
\begin{equation}\label{isergin}
t\left(\frac{1}{\lambda_{n,1}\lambda_{n,2}}\right)
t\left(\frac{q^2}{\lambda_{n,1}\lambda_{n,2}}\right) \;=\; \prod_k
q\mu_{k,1}\mu_{k,2}\left(1-q^2\frac{\lambda_{k,1}\lambda_{k,2}}{\lambda_{n,1}\lambda_{n,2}}\right)
\left(1-q^{-2}\frac{\lambda_{k,1}\lambda_{k,2}}{\lambda_{n,1}\lambda_{n,2}}\right)
\end{equation}
with $n=1,\dots, N$. These $N$ equations and conditions
(\ref{t0-tN}) do produce the whole spectrum of $t(u)$  for
inhomogeneous six-vertex model.

Equations (\ref{isergin}) may be obtained in the other way -- via
$\overline{Q}(v)=\langle v|\overline{Q}\rangle$. Condition
$\overline{Q}(0)=1$ fixes $u=u_n$,
\begin{equation}
u_n=\left(\lambda_{n,1}\lambda_{n,2}
q^{\Nop_{n,1}+\Nop_{n,2}}\right)^{-1}\;,
\end{equation}
corresponding $\overline{Q}=\overline{Q}_n(v)$ is a polynomial of
the power $\mathcal{J}_n=\Nop_{n,1}+\Nop_{n,2}$. In the six-vertex
case $\mathcal{J}_n=1$, therefore $\overline{Q}_n(v)=v-\zeta_n$,
and equating to zero each $v$-term of the right of equations
(\ref{T-Q}), one comes to
\begin{equation}
\zeta_n\;=\;-\frac{t(q^3u_n)}{\ds\prod_k
q\mu_{k,1}\mu_{k,2}(1-q^{-2}u_n/u_k)}\;=\;-\;
\frac{\ds\prod_k(1-q^2u_n/u_k)}{t(q u_n)}\;.
\end{equation}
The second equality gives (\ref{isergin}).

\subsection{$\mathcal{U}_q(\widehat{sl}_M)$ equations}

Let the lattice has arbitrary $N$ and $M$. We will consider
(\ref{T-Q}) for $\langle Q|u\rangle =Q(u)$ and for $\langle
u|\overline{Q}\rangle=\overline{Q}(u)$ -- the last function was
not mentioned before, but it is interesting to discuss what it is.

Equations $\langle Q|J(\uop,\vop)|u\rangle=0$ and $\langle
u|J(\uop,\vop)|\overline{Q}\rangle=0$ read correspondingly
\begin{equation}\label{TQ-slM}
\sum_{m=0}^M Q(q^{2m}u) (-v)^m
\Tau_m(u)\;=\;0\;\;\;\textrm{and}\;\;\; \sum_{m=0}^M \Tau(q^{-2m}
u) (-v)^m \overline{Q}(q^{-2m}u)\;=\;0\;,
\end{equation}
where
\begin{equation}
\Tau_m(u)\;=\;\sum_{n=0}^N (-q)^{nm} (-u)^{n} \mathbf{t}_{nm}\;,
\end{equation}
cf. (\ref{TslM},\ref{Tau}). Combinatorially, one can obtain the
following summation formulae: analogue of (\ref{t012})
\begin{equation}\label{u-per}
\begin{array}{l}
\ds \sum_n u^n t_{n,0} \;=\;
\prod_n\biggl(1+u\prod_m\lambda_{n,m}q^{\Nop_{n,m}}\biggr)\;,\\
\\
\ds \sum_n u^n t_{n,M} \;=\;
\prod_n\biggl(\prod_m\mu_{n,m}q^{\Nop_{n,m}} + u\prod_m
-q^{-1}\lambda_{n,m}\mu_{n,m}\biggr)\;,
\end{array}
\end{equation}
and analogue of (\ref{t0-tN})
\begin{equation}\label{v-per}
\begin{array}{l}
\ds \sum_m v^m t_{0,m} \;=\; \prod_m \biggl(1+v\prod_n \mu_{n,m}
q^{\Nop_{n,m}}\biggr)\;, \\ \\  \ds \sum_m v^m t_{N,m} \;=\;
\prod_m \biggl( \prod_n \lambda_{n,m} q^{\Nop_{n,m}} + v \prod_n
-q^{-1}\lambda_{n,m}\mu_{n,m}\biggr)\;.
\end{array}
\end{equation}
Let us fix notations for the inhomogeneities (cf. (\ref{un-vm}))
\begin{equation}\label{un-vm2}
u_n\;=\;\left(\prod_m
\lambda_{nm}q^{\Nop_{nm}}\right)^{-1}\;,\quad v_m=\left(\prod_n
\mu_{nm}q^{\Nop_{nm}}\right)^{-1}\;.
\end{equation}
It follows from (\ref{u-per}),
\begin{equation}\label{koncy}
\Tau_0(u)=\prod_n \left(1-\frac{u}{u_n}\right)\;,\;\;\;
\Tau_M(u)=\prod_{m} v_m \ \prod_n
\left(1-q^{-2\mathcal{J}_n}\frac{u}{u_n}\right)\;.
\end{equation}
For the normalization $Q(0)=\overline{Q}(0)=1$ both equations
(\ref{TQ-slM}) are equivalent to the first line of (\ref{v-per}):
\begin{equation}\label{S1}
\sum_{m=0}^M (-v)^m \Tau_m(0)=\prod_{m}
\left(1-\frac{v}{v_m}\right)=0
\end{equation}
Suppose, $Q(u)\sim u^K$ and $\overline{Q}(u)\sim u^{\overline{K}}$
when $u\to\infty$. Then (\ref{TQ-slM}) and the second line of
(\ref{v-per}) provides the following conditions for $K$ and
$\overline{K}$:
\begin{equation}\label{S2}
\prod_m \left(1-q^{2K-2\mathcal{K}_m}\frac{v}{v_m}\right)=0
\end{equation}
and
\begin{equation}\label{S3}
\prod_m\left(1-q^{-2\overline{K}-2N-2\mathcal{K}_m}\frac{v}{v_m}\right)=0\;.
\end{equation}
The charges $\mathcal{J}_n$ and $\mathcal{K}_m$ are given by
(\ref{charges}).  Equation (\ref{S1}) has the solutions $v=v_m$,
$m=1,2,3,\dots, M$. In what follows, we will assume the generic
set of $\mu_{nm}$, so that all $v_m$ are different. Let $v=v_m$
corresponds to $Q(u)=Q_m(u)$ and
$\overline{Q}(u)=\overline{Q}_m(u)$. Then (\ref{S2}) and
(\ref{S3}) defines the leading $u\to\infty$ asymptotic
\begin{equation}
Q_m(u)\sim u^{\mathcal{K}_m} \quad \textrm{and} \quad
\overline{Q}_m(u)\sim u^{\overline{\mathcal{K}}_m}\;,
\end{equation}
where
\begin{equation}\label{powers}
\overline{\mathcal{K}}_m\;=\;-\sum_n (1+\Nop_{n,m})\;.
\end{equation}\
Thus, for the Fock space representation all $Q_m(u)$ are
polynomials of the power $\mathcal{K}_m$. Functions
$\overline{Q}_m(u)$ are rational functions, below we construct
them in the terms of $Q_m(u)$.

Instead of the Fock space representation
$\textrm{Spec}(\Nop)=0,1,2,...,\infty$, one may consider the
anti-Fock space, $\textrm{Spec}(\Nop)=-1,-2,-3,...,-\infty$. Then
(\ref{local-inv-space}) are to be solved in different way, and as
the result $\overline{Q}_m(u)$ become polynomials and $Q_m(u)$
become rational functions.

The dual ``T-Q'' equations for $sl_N$ correspond to the evident
exchange $N\leftrightarrow M$, $u\leftrightarrow v$ and
$\mathcal{J}\leftrightarrow\mathcal{K}$.

Turn now to the form of $\overline{Q}_m$ for the Fock space
representation. The detailed ``arithmetical'' consideration of
(\ref{TQ-slM}) as a set of linear equations allows one to conclude
for instance
\begin{equation}\label{QQ1}
\overline{Q}_M(u)\;=\;\frac{W_M(u)}{V(u)}\;,
\end{equation}
where
\begin{equation}\label{QQ2}
W_M(u)\;=\;\det|| Q_j^{}(q^{2i}u) v_j^i||_{i,j=1,...,M-1}\;,
\end{equation}
and $\ds \frac{V(u)}{V(q^2u)} = v_1v_2...v_M
\frac{\Tau_M(q^Mu)}{\Tau_0(q^2u)}$. All the other $\overline{Q}_m$
correspond to (\ref{QQ1}) and (\ref{QQ2}) with permuted set of
indices of $Q_j$. Inhomogeneity of $v_m$ is important in this
consideration since if $Q_m(0)=1$, then $\ds W_M(0) =
v_1...v_{M-1} \prod_{1\leq i<j< M} (v_i-v_j) $, and therefore
$W_M(u)$ is not zero. As well, for the generic $v_m$ all $Q_m(u)$
are functionally independent, their Wronskian
\begin{equation}
W(u)\;=\;\det|| Q_j^{}(q^{2(i-1)}u)v_j^{i-1}||_{i,j=1,...,M}
\end{equation}
obeys
\begin{equation}
\frac{W(u)}{W(q^2u)}\;=\;v_1v_2...v_M\frac{\Tau_M(q^Mu)}{\Tau_0(u)}
\end{equation}
with the initial condition $\ds W(0)\;=\;\prod_{1\leq i<j\leq M}
(v_i-v_j)$, therefore $W(u)\neq 0$.

Using these ``arithmetical'' considerations, one may express the
fundamental transfer matrices $\Tau_m(u)$ of $sl_M$ via
determinants $\det||Q_j(q^{2p_i}u)v_j^{p_i}||_{i,j=1,...,M}$,
where $p_i$ is a subset of $(0,1,\dots. M)$. Moreover, for a more
general sets of $p_i$, any transfer matrix of $sl_M$ may be
expressed as such a determinant \cite{W3}.

The nested Bethe Ansatz equations (see \cite{Sutherland} and e.g.
\cite{deVega}) may be derived from the generalized ``T-Q''
equations in an ``arithmetical'' way as well. Let
\begin{equation}\label{nbae-Q}
\begin{array}{l}
\ds \mathcal{Q}_m(u) \;=\;\det || Q_j(q^{2i}u)
v_j^{i}||_{i,j=1,...,m}\;,\\
\ds \mathcal{P}_m(u) \;=\; \det || Q_j(q^{2i}u)
v_j^i||_{j=1,\dots,m; i=0,2,\dots,m}\;,
\end{array}
\end{equation}
with $\mathcal{Q}_0=1$ and $\mathcal{P}_0=0$. Function
$\mathcal{Q}_m(u)$ is a polynomial of the power
$\mathcal{K}_1+\mathcal{K}_2+\cdots+\mathcal{K}_m$. Equations
\begin{equation}\label{nbae-1}
\mathcal{P}_m(u)\mathcal{Q}_{m-1}(u) \;=\; \mathcal{P}_{m-1}(u)
\mathcal{Q}_m(u) + \frac{1}{v_m} \mathcal{Q}_{m-1}(q^2u)
\mathcal{Q}_m(q^{-2}u)
\end{equation}
for $m=1,2,\dots, M-1$ are just the determinant identities for
(\ref{nbae-Q}). For $m=M$ the system (\ref{nbae-1}) should be
completed by an equation, following from (\ref{TQ-slM}):
\begin{equation}\label{nbae-2}
\Tau_1(u) \mathcal{Q}_{M-1}(u) \;=\; \Tau_0(u)
\mathcal{P}_{M-1}(u) + v_1\cdots v_{M-1}\Tau_M(u)
\mathcal{Q}_{M-1}(q^2u)\;,
\end{equation}
where $\Tau_0$ and $\Tau_M$ are given by (\ref{koncy}). The nested
Bethe Ansatz equations are a closed system of algebraic equations
for roots of $\mathcal{Q}_m(u)$ following from
(\ref{nbae-1},\ref{nbae-2}).

\section{Conclusion}

This paper has a modest aim just a to give a correct form of
``T-Q'' equations. We can say nothing about their solution. But we
would like to note, that from the point of view of three
dimensional models, the thermodynamical limit is the limit $N,M\to
\infty$ with non-singular ratio $N:M$. The nested Bethe Ansatz
equations was never investigated in this limit since
$\mathcal{Q}_m(u)$ (\ref{nbae-Q}) with finite $m$ has a finite
number of roots. In addition, an excitation corresponds to a
change of the structure of occupation numbers. From $sl_M$ point
of view of the previous section, it corresponds not only to a
change of $\mathcal{K}_m$ related to the powers of
$\mathcal{Q}$-operators, but as well it corresponds to a change of
$\mathcal{J}_n$ which is a change of the $sl_M$-structure of the
nested Bethe Ansatz.

Let us better conclude this paper by a brief comparison of two
exactly integrable models in $2+1$ dimensional space-time. From
the point of view of the algebra of observables, these models
should be called ``$q$-oscillator model'' and ``Weyl-algebra
model''. The last one is a quantum-mechanical reformulation of
Zamolodchikov-Bazhanov-Baxter model of statistical mechanics,
which has a long history \cite{Z,Bax-W,Bax-Z,BB,MSS-vertex}. Both
models are based on two slightly different forms of local linear
problem, the linear problem for Weyl-algebra model may be found in
any of Refs. \cite{Sergeev-opus}. Solution of both classical
models may be expressed in terms of algebraic geometry,
\cite{Korepanov} and \cite{Sergeev-classic}. Equations of motion
may be understood as a canonical mapping conserving certain
symplectic structure, \cite{First} and \cite{Sergeev-symplectic}.
Poisson structure allows an immediate quantization, \cite{First}
and \cite{Sergeev-opus}. Quantum-mechanical integrals of motion
may be combined into a direct sum of transfer matrices for
fundamental representations of either $sl_N$ or $sl_M$,
\cite{First,letter} and \cite{Sergeev-transfermatrix}. And
finally, the solvability of the models is based on quantized
auxiliary linear problem and remarkable features of their
operator-valued matrices of coefficients, this paper and
\cite{Sergeev-lp}.

The continuous limit of both classical models may be illustrative.
Equations of motion for six fields $q_j^{},q_j^*$, $j=1,2,3$,
follow from the action
\begin{equation}
A\;=\;\int d^3x \left[  q_1^*\partial_1^{} q_1^{} +
q_2^*\partial_2^{} q_2^{} + q_3^*\partial_3^{} q_3^{} +
V(q^*,q)\right]\;,
\end{equation}
where for the classical continuous limit of $q$-oscillator model
$V=q_1^*q_2^*q_3^*-q_1^{}q_2^{}q_3^{}$, this is nothing but the
model of three-wave resonant interaction \cite{Kaup}. For the
classical continuous limit of Weyl-algebra model the potential is
$V=(q_1^*-q_2^{})(q_2^*-q_3^{})(q_3^*-q_1^{})$
\cite{Mangazeev-Sergeev}.

\noindent \textbf{Acknowledgements} The author should like to
thank V. Bazhanov, V. Mangazeev, M. Batchelor, M. Bortz and all
the Mathematical Physics group of the Department of Theoretical
Physics of RSPhysSE for fruitful discussions.


\begin{thebibliography}{99}

\bibitem{First}
V. Bazhanov and S. Sergeev, ``Zamolodchikov's Tetrahedron Equation
and Hidden Structure of Quantum Groups'', \emph{Preprint}
arXiv:het-th/0509181

\bibitem{letter}
S. Sergeev, ``Integrability of $q$-oscillator lattice model'',
\emph{Preprint} nlin.SI/0509043

\bibitem{Korepanov}
I. Korepanov, ``Algebraic integrable dynamical systems, $2+1$
dimensional models on wholly discrete space-time, and
inhomogeneous models on 2-dimensional statistical physics'',
\emph{Preprint} solv-int/9506003


\bibitem{bs}
V. Bazhanov and Yu. Stroganov, ``Conditions of commutativity of
transfer-matrices on a multidimensional lattice'', \emph{Theor.
Math. Phys.} \textbf{52} (1982) 685-691

\bibitem{W3}
V. V. Bazhanov, A. N. Hibberd and S. M. Khoroshkin ``Integrable
structure of $W_3$ Conformal Field Theory, Quantum Boussinesq
Theory and Boundary Affine Toda Theory'', \emph{Nucl. Phys.}
\textbf{B622} (2002) 475-547

\bibitem{Sutherland}
B. Sutherland, ``Model for a multicomponent quantum system'',
\emph{Phys. Rev. B} \textbf{12} (1975) 3795-3805

\bibitem{deVega}
H. J. de Vega, ``Yang-Baxter algebras, integrable theories and
Bethe ansatz. Proceedings of the Conference on Yang-Baxter
Equations'', \emph{Int. J. Mod. Phys. B} \textbf{4} (1990)
735--801

\bibitem{Z}
A. B.  Zamolodchikov, ``Tetrahedra equations and integrable
systems in three-dimensional space''. \emph{Soviet Phys. JETP}
\textbf{52} (1980) 325-336 [\emph{Zh. Eksp. Teor. Fiz.}
\textbf{79} (1980) 641--664]

A. B. Zamolodchikov, ``Tetrahedron equations and the relativistic
S matrix of straight strings in 2+1 dimensions'', \emph{Commun.
Math. Phys.} \textbf{79} (1981) 489-505

\bibitem{Bax-W}
R. J. Baxter, ``On Zamolodchikov's solution of the tetrahedron
equation'', \emph{Commun. Math. Phys.} \textbf{88} (1983) 185-205

\bibitem{Bax-Z}
R. J. Baxter, ``The Yang-Baxter Equations and the Zamolodchikov
Model'', \emph{Physica} \textbf{18D} (1986) 321-347

\bibitem{BB}
V. V. Bazhanov and R. J. Baxter, ``New solvable lattice models in
three dimensions'', \emph{J. Stat. Phys.} \textbf{69} (1992)
453-485

\bibitem{MSS-vertex}
S. M. Sergeev, V. V. Mangazeev and Yu. G. Stroganov, ``The vertex
reformulation of the Bazhanov-Baxter model'', \emph{J. Stat.
Phys.} \textbf{82} (1996) 31-50

\bibitem{Sergeev-opus}
S. Sergeev, ``Quantum {$2+1$} evolution model'', \emph{J. Phys. A:
Math. Gen.} \textbf{32} (1999) 5693--5714

S. Sergeev, ``Integrable three dimensional models in wholly
discrete space-time'', \emph{Integrable structures of exactly
solvable two-dimensional models of quantum field theory (Kiev,
2000)} NATO Sci. Ser. II Math. Phys. Chem. \textbf{35} (2001)
293--304 Kluwer Acad. Publ.

S. Sergeev, ``Complex of three-dimensional solvable models'',
\emph{J. Phys. A: Math. Gen.} \textbf{34} (2001) 10493--10503
(Symmetries and integrability of difference equations (Tokyo,
2000))

\bibitem{Sergeev-classic}
S. M. Sergeev, ``On exact solution of a classical 3{D} integrable
model'', \emph{J. Nonlinear Math. Phys.} \textbf{1} (2000) 57--72

\bibitem{Sergeev-symplectic}
S. M. Sergeev, ``$3{D}$ symplectic map'', \emph{Phys. Lett. A}
\textbf{253}  (1999) 145--150

\bibitem{Sergeev-transfermatrix}
S. M. Sergeev, ``Auxiliary transfer matrices for three-dimensional
integrable models'', \emph{Theoretical and Mathematical Physics}
\textbf{124} (2000) 391--409

\bibitem{Sergeev-lp}
S. M. Sergeev, ``Coefficient Matrices of a Quantum Discrete
Auxiliary Linear Problem'', \emph{Journal of Mathematical
Sciences} \textbf{115}(1) (2003) 2049-2057

\bibitem{Kaup}
D. J. Kaup, ``The inverse scattering solution for the full three
dimensional three-wave resonant interaction'', \emph{Physica}
\textbf{1D} (1980) 45-67


\bibitem{Mangazeev-Sergeev}
V. V. Mangazeev and S. M. Sergeev, ``The continuous limit of the
triple {$\tau$}-function model'', \emph{Theoretical and
Mathematical Physics} \textbf{129} (2001) 317--326


\end{thebibliography}
\end{document}